\documentstyle[preprint,aps]{revtex}

\def\pdif#1#2{\frac{\partial #1}{\partial #2}}
\def\dif#1#2{\frac{d #1}{d #2}}
\def\ppdif#1#2{\frac{\partial^2 #1}{\partial #2^2}}
\def\ddif#1#2{\frac{d^2 #1}{d #2^2}}

\begin{document}
\draft
\title{A new numerical scheme to compute 3D configurations 
of quasiequilibrium compact stars in general relativity \\  
-- Application to synchronously rotating binary star systems -- }
%
%
\author{Fumihiko Usui}
\address{Department of Earth Science and Astronomy,\\
         Graduate School of Arts and Sciences,\\
         University of Tokyo, Komaba, Meguro, Tokyo 153-8902, Japan}
\author{K\=oji Ury\=u}
\address{SISSA, Via Beirut 2-4, Trieste 34013, Italy}
\author{Yoshiharu Eriguchi}
\address{Department of Earth Science and Astronomy,\\
         Graduate School of Arts and Sciences,\\
         University of Tokyo, Komaba, Meguro, Tokyo 153-8902, Japan}

\date{\today}
\maketitle
\begin{abstract}
%
%

We have developed a new numerical scheme to obtain quasiequilibrium 
structures of nonaxisymmetric compact stars such as binary neutron star 
systems as well as the spacetime around those systems in general relativity.
Although, strictly speaking, there are no equilibrium states for binary
configurations in general relativity, the timescale of orbital change
due to gravitational wave radiation is so long compared with the orbital 
period that we can assume nonaxisymmetric systems in ``quasiequilibrium'' 
states.

Concerning quasiequilibrium states of binary systems in general relativity,
several investigations have been already carried out by assuming conformal
flatness of the spatial part of the metric.  However, the validity 
of the conformally flat treatment has not been fully 
analyzed except for axisymmetric configurations. Therefore it is desirable to 
solve quasiequilibrium states by developing totally different methods from 
the conformally flat scheme. In this paper we present a new numerical scheme 
to solve directly the Einstein equations for 3D configurations without 
assuming conformal flatness, although we make use of the simplified metric 
for the spacetime. This new formulation is the extension of the scheme which 
has been successfully applied for structures of axisymmetric rotating compact 
stars in general relativity. It is based on the integral representation of 
the Einstein equations by taking the boundary conditions at infinity into 
account. We have checked our numerical scheme by computing equilibrium 
sequences of binary polytropic star systems in Newtonian gravity and those of 
axisymmetric polytropic stars in general relativity.  We have applied this 
numerical code to binary star systems in general relativity and have 
succeeded in obtaining several equilibrium sequences of synchronously rotating 
binary polytropes with the polytropic indices $N = 0.0, 0.5$ and $1.0$.

\end{abstract}

\widetext

%
\section{Introduction}

One of the most important matters for theorists of relativistic astrophysics 
is to construct reliable models for quasiequilibrium configurations of binary 
neutron star systems. This is because such systems are the most promising 
sources of gravitational waves which can be observed by the gravitational 
wave detectors under construction, such as LIGO (USA), VIRGO (France/Italy), 
GEO (Germany/Great Britain), and TAMA (Japan). 

After the discovery of the first binary pulsar PSR1913+16 \cite{HT75}, 
the orbital change of PSR1913+16 revealed that the existence of gravitational 
wave did successfully explain the observational results \cite{TFM79}. From 
more detailed observations of PSR1913+16, much information about neutron stars 
such as the mass has been obtained and at the same time many other binary 
pulsar systems have been found \cite{TW82,TW89,TAMT93,ST85,AGKPW90}.
Since there are many binary systems composed of neutron stars/black holes, 
it is estimated that gravitational waves from binary systems will be observed 
directly once a year in galaxies within 200Mpc\cite{NPS91}.  Gravitational 
wave detectors will become new eyes in the 21st century for uncovering extreme 
states of the universe.

In order to understand coalescing stages of compact binary systems from
observation of gravitational waves, we need to know detailed processes of 
coalescing or merging phases. This can be done by performing dynamical 
computations of evolution of binary star systems to a very high accuracy. 
However, it is still difficult to do highly accurate numerical 
computations of 3D configurations in general relativity. One reason for that 
is the very existence of gravitational wave radiation from the systems.
There have been many attempts to take this approach and many
important results have been obtained (see, e.g., Oohara, Nakamura and 
Shibata\cite{ONS97}, Nakamura\cite{N98}), but fully satisfactory 
results have not been obtained yet. 

Recently several groups have begun to attack this problem from a different 
standpoint.  Although binary systems emit gravitational waves, the 
timescale of its effect on the orbital motion is rather long compared with
the orbital period except for the final stage of coalescence because the 
energy loss rate or the angular momentum loss rate due to gravitational 
waves is so small except for such a final phase. The order can be
estimated as $\displaystyle O\left((v/c)^5\right)$ where $v$ 
and $c$ are the typical velocity of the system and the velocity of light, 
respectively. Thus we can assume that the systems are in definite orbits.
If the binary systems are synchronously rotating, the systems can be 
approximately treated to be in equilibrium states by choosing frames 
which rotate with the same rotational periods as the orbital periods. 
In other words, we can treat the systems in `quasiequilibrium'.

From a different context, in Newtonian gravity, Hachisu and Eriguchi
\cite{HE84a,HE84b,H86,EH85} obtained first numerically
exact results for binary configurations of polytropes including incompressible 
models. In their papers, secular instability of synchronously rotating binary
systems and dynamical instability of asynchronously rotating binary systems
were discussed.  On the other hand, recent investigations in Newtonian gravity 
are deeply related to the problem mentioned above. Lai, Rasio and 
Shapiro\cite{LRS93a,LRS93b,LRS94a,LRS94b} have studied 
this problem by employing the ellipsoidal approximation scheme in which 
configurations are assumed to be exact ellipsoids. More recently Uryu and 
Eriguchi\cite{UE98a,UE98b,UE98c,UE99} have developed a scheme for
irrotational binary star systems and obtained numerically exact stationary 
sequences. Newly obtained stationary sequences have been analyzed and much 
information about dynamical instability of evolutionary sequences of 
polytropic binary star systems has been obtained.

However, Newtonian configurations cannot be directly applied to realistic 
evolution for binary compact star systems. Thus some authors have studied 
binary configurations in post-Newtonian regime
\cite{AS96,S96,LRS97,S97,TS97,ST97,TAS98,T99}.  
Post-Newtonian analyses show that the critical angular velocity for 
instability is increased by 10-15 \% compared with that of Newtonian gravity. 
It is necessary to solve general relativistic models for binary 
configurations to compute exact critical states against instabilities.

The first numerical results for quasiequilibrium states of binary 
configurations in full general relativity have been obtained by Wilson, 
Mathews and Marronetti\cite{WMM96}. In order to treat the problem tractable, 
they have assumed that the spatial part of the metric is 
{\it conformally flat} (the conformally flat condition (CFC), hereafter).  
One of their results was surprising: binary neutron stars 
become unstable and collapse into black holes individually prior to merging
(see also Mathews and Wilson\cite{MW97}). The same problem has been 
investigated by applying almost the same formulation but by using a different 
numerical scheme by Baumgarte et al.\cite{BCSST98a} (also see Cook, Shapiro
and Teukolsky\cite{CST96}, Baumgarte et al.\cite{BCSST98b}). Baumgarte
et al.\cite{BCSST98a} have found that there exist quasiequilibrium sequences 
of binary systems just prior to merger, i.e. that individual
collapse to black holes will not occur (see also \cite{RS99,BGM99}).

In order to check the validity of the CFC and its accuracy, Cook, Shapiro 
and Teukolsky\cite{CST96} have computed equilibrium configurations
of axisymmetric rotating polytropes by using two different schemes:
1) the scheme with the CFC and 2) the KEH scheme\cite{KEH89}. 
Since differences of physical quantities between two schemes are less 
than 5 \%, they have concluded that the scheme with the CFC will give 
reasonably accurate results even for other situations such as binary 
configurations.  However, its validity is not fully understood \cite{AS96}, 
for example, it is uncertain whether results from axisymmetric configurations 
may be applied to nonaxisymmetric situations or not.  In this sense, 
it is desirable to develop different schemes from that with the CFC and to 
compare results of two or many different schemes for nonaxisymmetric models.

In this paper, we present a new numerical scheme to handle 
quasiequilibrium states of nonaxisymmetric configurations in general 
relativity. This scheme is an extension of the numerical scheme for
axisymmetric configurations developed by Komatsu, Eriguchi and Hachisu
(KEH scheme)\cite{KEH89}.  The basic idea used in \cite{KEH89} is to transform
the Einstein equations into integral equations by using the Green function
for the Laplacian in the flat space. Since we can include the boundary 
conditions into the Green function, what we need to do is to solve integral
equations by some means.  In order to adopt the same procedure for 
nonaxisymmetric models as that for axisymmetric configurations, the crucial 
step is to make up the Laplacian in the flat space by arranging the Einstein 
equations.  For nonaxisymmetric configurations, however, the Einstein 
equations for each metric component do not contain Laplacians in the flat 
space even after proper arrangements.  This can be done if we add appropriate 
terms to both sides of properly selected component equations of the Einstein 
equations, although the source terms of those equations contain extra 
derivatives in addition to the original sources.  

As the first step to construct realistic models for binary star systems, 
we have assumed a simplified form for the metric.  The basic equations 
consist of combinations of dominant components of the Einstein equations 
for this simplified metric.  We have successfully applied the above procedure 
to make up Laplacians in the flat space and have succeeded in developing
a new reliable numerical scheme.  

This paper is organized as follows.  In section II, we introduce assumptions 
and basic formulation of the problem.  The choice of the simplified metric 
and the Einstein equations are explained.  In section III,  we describe 
the numerical solving method briefly.  In section IV, results of our
numerical computations are presented.  We have tested our new method by 
comparing our results with (1) those of the axisymmetric configurations 
of fully relativistic and rapidly rotating polytropes and  (2) those of
the Newtonian binary systems.  In section V, we discuss the validity of the 
present scheme and the future prospect.

\section{Assumptions and basic equations}
\subsection{Assumptions}

As mentioned in Introduction, nonaxisymmetric rotating objects
cannot be in equilibrium states in the framework of general relativity.  
It is because gravitational waves carry away the energy as well as the 
angular momentum from the binary star systems to infinity and so no conserved 
physical quantities can exist for them. However, for almost all stages except 
for the last few milliseconds of coalescence of binary systems, the timescale 
of change of the system due to gravitational waves, $\tau_{\rm GW}$, is much 
longer than that of the orbital period, $\tau_{\rm orb}$. Specifically,
$\tau_{\rm orb}/\tau_{\rm GW}\sim (v/c)^5$ and it is less than 1 \% even for
$v\sim 0.3c$.  At the same time, the energy radiated due to gravitational
wave emission is only a few percent of the total energy even when two stars 
come very close. Therefore, gravitational waves can be neglected for most 
stages of evolution of binary star systems. 

If we neglect the effect of gravitational waves, we can choose a proper
rotating frame in which the system is regarded as a stationary one. If we let
the angular velocity of this frame seen from the observer at infinity be
$\Omega$, the following vector in the inertial frame:
\begin{equation}
\vec{\xi} = \pdif{}{t} + \Omega \pdif{}{\varphi} \ ,
\end{equation}
can be regarded as a Killing vector (see e.g. Bonazzola, Frieben and
Gourgoulhon\cite{BFG96,BFG98}, Bonazzola, Gourgoulhon and 
Marck\cite{BGM97}), where $t$ and $\varphi$ are the time and the azimuthal 
coordinates, respectively. Thus we assume that nonaxisymmetric systems are
in quasiequilibrium states.

Assumptions we will make in this paper are as follows:
\begin{description}
\item [1)] We will treat a binary system which consists of two equal mass
stars in a circular orbit, although our code can treat a single 
nonaxisymmetric body.
\item [2)] The binary star system is assumed to be in a stationary state
in the rotating frame with the angular velocity $\Omega$. In other words, 
we will neglect gravitational waves from the binary star system.
\item [3)] Axes of spins of two stars and that of the orbital motion are 
parallel to each other.  A schematic figure of the system is shown in Figure
\ref{coordinate system}.  The $z-$axis is the rotational axis of the orbital 
motion.  The $x-$axis is connecting two centers of mass of the stars. The 
intersection of these two axes is defined as the origin of the system. 
The $y-$axis is perpendicular to the $x-$ and $z-$axes. We will call the 
$x-y$ plane as the equatorial plane.
\item [4)] Spins of two stars are synchronized to the orbital motion.
Each star is rigidly rotating with the angular velocity
$\Omega$ if seen from a distant place. 
\item [5)] The matter of the star is perfect fluid and a polytropic relation
is assumed.
\item [6)] The matter distribution and the spacetime are assumed to be
symmetric about three planes: the equatorial plane, the $y-z$ plane
and the $x-z$ plane.
\end{description}

\subsection{Metric and Einstein equations}
In addition to the assumptions mentioned above, we further assume
the following form for the metric in the spherical coordinates 
$(r, \theta, \varphi)$ (hereafter we use the units of $c=G=1$):
\begin{equation}
\label{metric}
ds^2  = - e^{2\nu}dt^2
     	 + r^2\sin^2\theta e^{2\beta}(d\varphi -\omega dt)^2
         + e^{2\alpha}dr^2  
	 + r^2 e^{2\alpha'}d\theta^2 \ , 
\end{equation}
where $\nu$, $\beta$, $\omega$, $\alpha$ and $\alpha'$ are the metric
coefficients and they are functions of $r$, $\theta$ and $\varphi$.  
This choice of the metric is not the most general one for quasiequilibrium 
states because we do not take into account $t r$- and/or $t \theta$- 
components of the metric. However, since the purpose of this paper is to show 
the effectiveness of our new scheme and the $t \varphi$- component is 
considered to be the most dominant one among nondiagonal components, 
we use the above form for the metric in this paper.  Furthermore, for 
simplicity, we assume the following condition:
\begin{equation}
\alpha = \alpha' \ .
\end{equation}
This form of the metric becomes exact for stationarily axisymmetric 
configurations. Although the assumed metric form is incomplete for
quasiequilibrium 3D configurations, the essential and technical part of a 
new numerical scheme for 3D configurations can be shown by choosing this 
kind of simplified metric and will be extended to more general form of
the metric.

As mentioned before, the equation of state for the matter is assumed to be 
polytropic as follows:
\begin{equation}
\label{polytrope}
p = K \varepsilon ^{1+1/N} \ ,
\end{equation}
where $p$, $K$, $\varepsilon$ and $N$ are the pressure, a constant, the energy
density and the polytropic index, respectively.  It should be noted that
this polytropic relation is slightly different from that used in 
\cite{BCSST98a}.  This equation of state can be rewritten as follows 
by using the Lane-Emden function $\lambda$:
\begin{eqnarray}
p &=& p_{c}\lambda^{1+N} \ ,\\
\varepsilon &=& \varepsilon_{c}\lambda^N \ ,
\end{eqnarray}
where $p_c$ and $\varepsilon_c$ are the maximum pressure and the maximum 
energy density, respectively.

The energy-momentum tensor, $T^{\zeta \eta}$, for the perfect fluid is written
as:
\begin{equation}
T^{\zeta \eta} = ( \varepsilon + p )u^{\zeta} u^{\eta} + p g^{\zeta \eta} \ ,
\end{equation}
where $u^{\zeta}$ and $g^{\zeta \eta}$ are the four velocity of the matter and
the metric, respectively.
Throughout this paper, Greek indices run from 0 to 3.
Concerning the four velocity, since an observer in the rotating frame sees
that the matter is static, we can obtain the following relation by using
the condition $u^{\zeta} u_{\zeta} = -1$:
\begin{equation}
u^{\zeta} = \frac{e^{-\nu}}{\sqrt{1-v^2}} \pmatrix{1 \cr 0 \cr 0 \cr 0 \cr} \ ,
\end{equation}
where $v$ is the proper velocity of the matter:
\begin{equation}
v =  r ( \Omega - \omega ) \sin\theta e^{\beta-\nu} \ .
\end{equation}

After a lengthy but straightforward calculation of the Ricci tensor from the 
metric (\ref{metric}) (see Appendix), combining the obtained Einstein 
equations appropriately (see Appendix) and introducing the following two 
variables $\rho$ and $\gamma$ by
\begin{eqnarray}
\rho \equiv \beta - \nu \ ,\\
\gamma \equiv \beta + \nu \ ,
\end{eqnarray}
we can obtain the following equations for the metric functions
$\rho, \gamma, \alpha$ and $\omega$:

\def\pdif#1#2{\frac{\partial #1}{\partial #2}}
\def\dif#1#2{\frac{d #1}{d #2}}
\def\ppdif#1#2{\frac{\partial^2 #1}{\partial #2^2}}
\def\ddif#1#2{\frac{d^2 #1}{d #2^2}}

\def\gam{\gamma}
\def\alp{\alpha}
\def\omg{\omega}

\def\drhop{\pdif{\rho}{\varphi}}
\def\dgamp{\pdif{\gam}{\varphi}}
\def\dalpp{\pdif{\alp}{\varphi}}
\def\domgp{\pdif{\omg}{\varphi}}
\def\dbetap{\pdif{\beta}{\varphi}}
\def\dnup{\pdif{\nu}{\varphi}}

\def\ddrho{\ppdif{\rho}{\varphi}}
\def\ddgam{\ppdif{\gam}{\varphi}}
\def\ddalp{\ppdif{\alp}{\varphi}}
\def\ddomg{\ppdif{\omg}{\varphi}}
\def\ddbeta{\ppdif{\beta}{\varphi}}
\def\ddnu{\ppdif{\nu}{\varphi}}

\def\rrss{r^2\sin^2\theta}

\begin{eqnarray}
\label{rho}
\Delta(\rho e^{\gam/2}) &=& S_{\rho}(r,\theta,\varphi) \ ,\\
\label{gamma}
\Delta(\gam e^{\gam/2}) &=& S_{\gam}(r,\theta,\varphi) \ ,\\
\label{alpha}
\Delta\alp &=& S_{\alp}(r,\theta,\varphi) \ ,\\
\label{omega}
(\Delta+\frac{2}{r}\pdif{}{r}+\frac{2\cot\theta}{r^2}\pdif{}{\theta})(\omg e^{\rho+\gam/2}) &=& S_{\omg}(r,\theta,\varphi) \ ,
\end{eqnarray}
where
\begin{eqnarray}
\Delta \equiv \ppdif{}{r}+\frac{2}{r}\pdif{}{r}+\frac{1}{r^2}\ppdif{}{\theta}+\frac{1}{r^2}\cot\theta\pdif{}{\theta}+\frac{1}{r^2\sin^2\theta}\ppdif{}{\varphi} \ ,
\end{eqnarray}
and
\begin{eqnarray}
S_{\rho}(r,\theta,\varphi) &=&
e^{\gam/2}\left[-8\pi e^{2\alp}(\varepsilon +p)\frac{1+v^2}{1-v^2} - \left(\frac{1}{r}\pdif{\gam}{r}+\frac{1}{r^2}\cot\theta\pdif{\gam}{\theta}\right)\right.\nonumber\\
& &{}\hspace{1cm}-2\frac{e^{2\alp-2\beta}}{\rrss}\left(\ddalp+(\dalpp)^2-\dgamp\dalpp\right)\nonumber\\
& &{}\hspace{1cm}-2(\Omega-\omega)^2e^{2\alp-2\nu}\left(\ddalp+(\dalpp)^2-\dgamp\dalpp\right)\nonumber\\
& &{}\hspace{1cm}-\rrss e^{2\rho}\left((\pdif{\omega}{r})^2+\frac{1}{r^2}(\pdif{\omega}{\theta})^2\right)\nonumber\\
& &{}\hspace{1cm}+\frac{\rho}{2}\left\{16\pi e^{2\alpha}p - \left(\frac{1}{r}\pdif{\gam}{r}+\frac{1}{r^2}\cot\theta\pdif{\gam}{\theta}\right)-\frac{1}{2}\left((\pdif{\gam}{r})^2+\frac{1}{r^2}(\pdif{\gam}{\theta})^2\right)\right.\nonumber\\
& &{}\hspace{2cm}-2\frac{e^{2\alp-2\beta}}{\rrss}\left(\ddnu+(\dnup)^2+\ddalp+(\dalpp)^2-\drhop\dalpp-\dnup\dbetap\right)\nonumber\\
& &{}\hspace{2cm}+2(\Omega-\omega)^2e^{2\alp-2\nu}\left(\ddbeta+(\dbetap)^2+\ddalp+(\dalpp)^2+\drhop\dalpp-\dnup\dbetap\right)\nonumber\\
& &{}\hspace{2cm}+2(\Omega-\omega)e^{2\alp-2\nu}\left(-\ddomg-\domgp(2\dalpp+2\drhop+\dgamp)\right)\nonumber\\
& &{}\hspace{2cm}\left.\left.+2e^{2\alp-2\nu}(\domgp)^2\right\}\right]\nonumber\\
& &{}+\frac{1}{\rrss}\ppdif{}{\varphi}(\rho e^{\gam/2}) \ ,
\end{eqnarray}

\begin{eqnarray}
S_{\gam}(r,\theta,\varphi) &=&
e^{\gam/2}\left[16\pi e^{2\alp}p-\left(\frac{1}{r}\pdif{\gam}{r}+\frac{1}{r^2}\cot\theta\pdif{\gam}{\theta}\right)\right.\nonumber\\
& &{}\hspace{1cm}+\frac{\gam}{2}\left\{16\pi e^{2\alp}p-\left(\frac{1}{r}\pdif{\gam}{r}+\frac{1}{r^2}\cot\theta\pdif{\gam}{\theta}\right)-\frac{1}{2}\left((\pdif{\gam}{r})^2+\frac{1}{r^2}(\pdif{\gam}{\theta})^2\right)\right.\nonumber\\
& &{}\hspace{2cm}-2\frac{e^{2\alp-2\beta}}{\rrss}\left(\ddnu+(\dnup)^2+\ddalp+(\dalpp)^2-\drhop\dalpp-\dnup\dbetap\right)\nonumber\\
& &{}\hspace{2cm}+2(\Omega-\omega)^2e^{2\alp-2\nu}\left(\ddbeta+(\dbetap)^2+\ddalp+(\dalpp)^2+\drhop\dalpp-\dnup\dbetap\right)\nonumber\\
& &{}\hspace{2cm}+2(\Omega-\omega)e^{2\alp-2\nu}\left(-\ddomg-\domgp(2\dalpp+2\drhop+\dgamp)\right)\nonumber\\
& &{}\hspace{2cm}\left.+2e^{2\alp-2\nu}(\domgp)^2\right\}\nonumber\\
& &{}\hspace{1cm}-2\frac{e^{2\alp-2\beta}}{\rrss}\left(\ddnu+(\dnup)^2+\ddalp+(\dalpp)^2-\drhop\dalpp-\dnup\dbetap\right)\nonumber\\
& &{}\hspace{1cm}+2(\Omega-\omega)^2e^{2\alp-2\nu}\left(\ddbeta+(\dbetap)^2+\ddalp+(\dalpp)^2+\drhop\dalpp-\dnup\dbetap\right)\nonumber\\
& &{}\hspace{1cm}+2(\Omega-\omega)e^{2\alp-2\nu}\left(-\ddomg-\domgp(2\dalpp+2\drhop+\dgamp)\right)\nonumber\\
& &{}\hspace{1cm}\left.+2e^{2\alp-2\nu}(\domgp)^2\right]\nonumber\\
& &{}+\frac{1}{\rrss}\ppdif{}{\varphi}(\gam e^{\gam/2}) \ ,
\end{eqnarray}

\begin{eqnarray}
S_{\alp}(r,\theta,\varphi) &=&
-4\pi e^{2\alp}(\varepsilon+p)+\frac{1}{r}\pdif{\alp}{r}+\frac{1}{r^2}\cot\theta\pdif{\alp}{\theta}+\frac{1}{r}\pdif{\nu}{r}+\frac{1}{r^2}\cot\theta\pdif{\nu}{\theta}\nonumber\\
& &{}+\pdif{\nu}{r}\pdif{\beta}{r}+\frac{1}{r^2}\pdif{\nu}{\theta}\pdif{\beta}{\theta}\nonumber\\
& &{}+\frac{1}{4}\rrss e^{2\rho}\left((\pdif{\omg}{r})^2+\frac{1}{r^2}(\pdif{\omg}{\theta})^2\right)\nonumber\\
& &{}+\frac{e^{2\alp-2\beta}}{\rrss}\left(\ddnu+(\dnup)^2-\dnup\dbetap-(\dalpp)^2\right)\nonumber\\
& &{}-(\Omega-\omega)^2e^{2\alp-2\nu}\left(\ddbeta+(\dbetap)^2-\dnup\dbetap-(\dalpp)^2\right)\nonumber\\
& &{}-(\Omega-\omega)e^{2\alp-2\nu}\left(-\ddomg-\domgp(2\drhop+\dgamp)\right)-e^{2\alp-2\nu}(\domgp)^2\nonumber\\
& &{}+\frac{1}{\rrss}\ddalp \ ,
\end{eqnarray}

\begin{eqnarray}
S_{\omg}(r,\theta,\varphi) &=&
e^{\rho+\gam/2}\left[-16\pi e^{2\alp}\frac{(\varepsilon+p)(\Omega-\omega)}{1-v^2}\right.\nonumber\\
& &{}\hspace{1cm}+\omega\left\{-8\pi e^{2\alp}\frac{(1+v^2)\varepsilon+2v^2p}{1-v^2}\right.\nonumber\\
& &{}\hspace{2cm}+\frac{1}{r}\left(2\pdif{\rho}{r}-\frac{1}{2}\pdif{\gam}{r}\right)+\frac{1}{r^2}\cot\theta\left(2\pdif{\rho}{\theta}-\frac{1}{2}\pdif{\gam}{\theta}\right)\nonumber\\
& &{}\hspace{2cm}+\left((\pdif{\rho}{r})^2-\frac{1}{4}(\pdif{\gam}{r})^2\right)+\frac{1}{4r^2}\left(4(\pdif{\rho}{\theta})^2-(\pdif{\gam}{\theta})^2\right)\nonumber\\
& &{}\hspace{2cm}-\frac{e^{2\alp-2\beta}}{\rrss}\left(\ddnu+(\dnup)^2+3(\ddalp+(\dalpp)^2)-\dalpp(\drhop+2\dnup)-\dnup\dbetap\right)\nonumber\\
& &{}\hspace{2cm}+(\Omega-\omega)^2e^{2\alp-2\nu}\left(\ddbeta+(\dbetap)^2-(\ddalp+(\dalpp)^2)\right.\nonumber\\
& &{}\hspace{2cm}\left.+(\drhop+2\dnup)\dalpp-\dnup\dbetap\right)\nonumber\\
& &{}\hspace{2cm}+(\Omega-\omega)e^{2\alp-2\nu}\left(-\ddomg-\domgp(2\dalpp+2\drhop+\dgamp)\right)+e^{2\alp-2\nu}(\domgp)^2\nonumber\\
& &{}\hspace{2cm}\left.-\rrss e^{2\rho}\left((\pdif{\omg}{r})^2+\frac{1}{r^2}(\pdif{\omg}{\theta})^2\right)\right\}\nonumber\\
& &{}\hspace{1cm}\left.-4\frac{e^{2\alp-2\beta}}{\rrss}(\Omega-\omega)\left(\ddalp+(\dalpp)^2-\dalpp\dgamp\right)\right]\nonumber\\
& &{}+\frac{1}{\rrss}\ppdif{}{\varphi}(\omega e^{\rho+\gam/2}) \ .
\end{eqnarray}

If we multiply ($r\sin\theta\sin\varphi$) to both sides of Equation 
(\ref{omega}), we can obtain the following equation:

\begin{equation}
\label{omega new}
\Delta(r\sin\theta\sin\varphi \cdot \omega e^{\rho+\gam/2}) = \tilde{S}_{\omega}(r,\theta,\varphi) \ ,
\end{equation}
where
\begin{equation}
\tilde{S}_{\omega}(r,\theta,\varphi) = r\sin\theta\sin\varphi S_{\omega}(r,\theta,\varphi) + \frac{2\cos\varphi}{r\sin\theta}\pdif{}{\varphi}(\omega e^{\rho+\gamma/2}) \ .
\end{equation}

It should be noted that in these equations we make up Laplacians in
the flat space by adding terms with derivatives 
$\displaystyle \frac{1}{r^2\sin^2\theta}\ppdif{}{\varphi}$
to both sides of the equations. The rest of the Einstein equations are not 
used in this paper.  Therefore, our solutions are not numerically exact ones 
which satisfy all components of the Einstein equations. This is similar to 
the CFC approach \cite{WMM96,BCSST98a}.  We will discuss the choice of 
the components of the Einstein equations in a later section.

\subsection{Boundary conditions and integral representation of the Einstein 
equations}
Equations (\ref{rho}) -- (\ref{alpha}) and (\ref{omega new}) can be 
regarded as Poisson equations for the corresponding quantities
if we treat the right hand sides as source terms, although source
terms contain unknown quantities. 

Under our assumption of quasiequilibrium states or no emission of
gravitational waves, we need not consider that there are gravitational 
waves at infinity.  This implies that the metric at infinity must be flat. 
Therefore,  we can transform these equations by using the Green
function for the Laplacian in the flat space, $1/|{\bf r}-{\bf r'}|$, into
the integral equations as follows:
\begin{eqnarray}
\label{int rho}
\rho &=& -\frac{1}{4\pi}e^{-\gamma/2}\int_{0}^{\infty}dr'\int_{0}^{\pi}d\theta'\int_{0}^{2\pi}d\varphi'\sin\theta'r'^2S_{\rho}(r',\theta',\varphi')\frac{1}{|{\bf r}-{\bf r^{\prime}}|} \ ,\\
\label{int gamma}
\gamma &=& -\frac{1}{4\pi}e^{-\gamma/2}\int_{0}^{\infty}dr'\int_{0}^{\pi}d\theta'\int_{0}^{2\pi}d\varphi'\sin\theta'r'^2S_{\gamma}(r',\theta',\varphi')\frac{1}{|{\bf r}-{\bf r^{\prime}}|} \ ,\\
\label{int alpha}
\alpha &=& -\frac{1}{4\pi}\int_{0}^{\infty}dr'\int_{0}^{\pi}d\theta'\int_{0}^{2\pi}d\varphi'\sin\theta'r'^2S_{\alpha}(r',\theta',\varphi')\frac{1}{|{\bf r}-{\bf r^{\prime}}|} \ ,\\
\label{int omega}
r\sin\theta\sin\varphi \omega &=& -\frac{1}{4\pi}e^{-(\rho+\gamma/2)}\int_{0}^{\infty}dr'\int_{0}^{\pi}d\theta'\int_{0}^{2\pi}d\varphi'\sin\theta'r'^2\tilde{S}_{\omega}(r',\theta',\varphi')\frac{1}{|{\bf r}-{\bf r^{\prime}}|} \ .
\end{eqnarray}
The Green function in these equations is expanded as follows:
\begin{eqnarray}
\label{green}
\frac{1}{ |{\bf r}-{\bf r^{\prime}}|}  &=&  \sum^\infty_{n=0}f_n(r,r')
          \left[ P_n(\cos\theta)P_n(\cos\theta') \right. \nonumber \\
          & & \left. + 2\sum^n_{m=1}\frac{(n-m)!}{(n+m)!}
           P_n^m(\cos\theta)P_n^m(\cos\theta')
       \cos m(\varphi-\varphi') \right] \ ,
\end{eqnarray}
where
\begin{eqnarray}
f_n(r,r') = \cases{\displaystyle \frac{1}{r}\left(\frac{r'}{r}\right)^n \ ,
& {\rm for} $r'/r \leq 1$ \cr
                   \displaystyle \frac{1}{r'}\left(\frac{r}{r'}\right)^n \ ,
& {\rm for} $r/r' > 1$   \cr}
\end{eqnarray}
and $P_n$ and $P_n^m$ are the Legendre polynomials and the associated Legendre
functions, respectively. The asymptotic flatness at infinity, 
$r \rightarrow \infty$, 
\begin{eqnarray}
\label{rho inf}
\rho &\sim& O( 1/r ) \ , \\
\label{gamma inf}
\gamma &\sim& O( 1/r^2 ) \ , \\
\label{alpha inf}
\alpha &\sim& O( 1/r ) \ , \\
\label{omega inf}
\omega &\sim& O( 1/r^3 ) \ , 
\end{eqnarray}
are satisfied automatically, if the source terms behave properly.

\subsection{Hydrostatic equation}
The hydrostatic equation is derived from the conservation law 
$T^{\zeta \eta}_{ \ \  ;\eta} = 0$ as follows:
\begin{equation}
\nabla p - (\varepsilon + p) \nabla \ln u^t = 0 \ .
\end{equation}
By using the polytropic equation (\ref{polytrope}) and the rigid rotation law,
it can be integrated to the following equation:
\begin{equation}
\label{hydro}
(1+N)\ln \left(K \varepsilon^{1/N} + 1\right) + \nu + \frac{1}{2} \ln(1-v^2)  = C \ ,
\end{equation}
where $C$ is a constant of integration.

\section{Method of solution}
\subsection{Model parameters}
Once the equation of state is fixed, we need to specify two parameters
to determine one model for a rotating equilibrium configuration: one parameter 
which represents the strength of gravity and the other for the amount of 
rotation.  For polytropes, we have to specify one more parameter because
the constant $K$ is a free parameter.  In our formulation, we choose the
following three parameters: 1) the maximum energy density, $\varepsilon_c$,
2) the ratio of the maximum pressure to the maximum energy density, $\kappa$:
\begin{equation}
\kappa \equiv \frac{p_c}{\varepsilon_c} \ ,
\end{equation}
and 3) the ratio of the shortest distance $r_A$ (distance from the origin to 
point A, see Figure \ref{coordinate system}) to the largest distance 
$r_B$ (distance from the origin to point B, see Figure 
\ref{coordinate system}), $q$:
\begin{equation}
q \equiv \frac{r_A}{r_B} \ .
\end{equation}
The value of the quantity $\kappa$ represents the strength of gravity because
$\displaystyle \kappa \sim \frac{r_g}{R}$, where $r_g$ and $R$ are the 
Schwarzschild radius and the radius of the star, respectively. This is regarded
as the parameter of compactness.For post-Newtonian models, $\kappa \leq 0.1$ and
for typical neutron stars, $\kappa \sim 0.2 - 0.4$.  

The quantity $q$ indirectly specifies the rotation rate.  It may be possible 
to choose alternatively the angular velocity or the angular momentum to 
specify the rotation rate.  However, from the numerical computational point
of view, it is much better to choose the ratio of two distances instead
of choosing the other physical quantities \cite{H86}.

For polytropes, we can introduce dimensionless physical quantities by using
two constants $c$ and $G$ as well as the maximum energy density 
$\varepsilon_c$. Since the quantity $\varepsilon_c$ does not appear
in the basic equations,  we can treat the problem
with two parameters, $\kappa$ and $q$, in addition to the polytropic
index $N$. In summary, once we specify these parameters, all we have to do 
is to solve for the metric coefficients $\rho$, $\gamma$, $\alpha$ and 
$\omega$, distributions of the energy density $\varepsilon$, the pressure 
$p$ and the angular velocity of the system $\Omega$.

In actual numerical computations, we further normalize the radial coordinate 
and the metric functions as follows:
\begin{equation}
r = r_0 \sqrt{f_x} \tilde{r} \ ,
\end{equation}
and 
\begin{eqnarray}
\rho = f_x \tilde{\rho} \ , \\
\gamma = f_x \tilde{\gamma} \ , \\
\alpha = f_x \tilde{\alpha} \ , \\
\omega = \Omega_0 f_x \tilde{\omega} \ , 
\end{eqnarray}
where 
\begin{eqnarray}
r_0 \equiv \frac{c^2}{\sqrt{G \varepsilon_c}} \ , \\
\Omega_0 \equiv \sqrt{4 \pi G \varepsilon_c/c^2} \ , 
\end{eqnarray} 
and $f_x$ is a dimensionless normalization constant which is implicitly 
determined from
\begin{equation}
r_B = \sqrt{f_x} r_0 \ , \ {\rm or} \ \ \tilde{r}_B = 1 \ .
\end{equation}
By using these normalized quantities, the hydrostatic equation (\ref{hydro}) 
is rewritten as:
\begin{equation}
\label{hydro new}
(1+N)\ln \left(\kappa \lambda + 1\right) + f_x \tilde{\nu} + 
\frac{1}{2} \ln(1-v^2)  = C \ ,
\end{equation}
where
\begin{equation}
v =  \sqrt{f_x} \tilde{r} ( \tilde{\Omega} - f_x \tilde{\omega} ) 
\sin\theta\exp\{{f_x(\tilde{\beta}-\tilde{\nu})\}} \ .
\end{equation}
Here
\begin{equation}
\tilde{\Omega} \equiv \frac{\Omega}{\Omega_0} \ .
\end{equation}

\subsection{Solving scheme}
Our formulation in this paper is almost the same as that used in Komatsu,
Eriguchi and Hachisu\cite{KEH89}.  Therefore we can adopt the HSCF 
method\cite{H86} or the KEH method\cite{KEH89} as our solving scheme. 
The HSCF method is the extended version of the the Self-Consistent Field
method developed by Ostriker and Mark \cite{OM68}. 

The essence of the HSCF method or the KEH method can be briefly summarized as 
follows: At the beginning of the computation, we prepare initial guesses for 
the metric potentials $\rho$, $\gamma$, $\alpha$ and $\omega$, the energy 
density $\varepsilon$, and the angular velocity $\Omega$ as well as
the quantity $f_x$.  Substituting them into right hand sides of the 
integral equations (\ref{int rho})-(\ref{int omega}), we obtain new values 
of $\rho$, $\gamma$, $\alpha$ and $\omega$.  At this point, 
we must solve the angular velocity $\Omega$, a constant of integration $C$ 
and the scale parameter $f_x$. These quantities are obtained from the 
hydrostatic equations (\ref{hydro new}) at points A, B (in Figure 
\ref{coordinate system}) and at point of the maximum density.  Using newly 
obtained $\rho$, $\gamma$, $\alpha$, $\omega$, $\Omega$, $C$ and $f_x$,
we calculate the new energy density $\varepsilon$ from the hydrostatic 
equation (\ref{hydro new}).  

These newly obtained values of $\rho$, $\gamma$, $\alpha$, $\omega$, 
$\varepsilon$, $\Omega$ and $f_x$ are used as the trial guesses in the next 
iteration cycle. We repeat this procedure until the relative differences 
between new values and old ones become small enough, i.e., typically
less than $10^{-4}$.  At this point we regard these values as converged
ones.

This iteration cycle is carried out by fixing $N$, $\kappa$ and $q$.
After we obtain one equilibrium model, we solve other models by changing the 
axis ratio $q$.  In this way we can obtain a sequence of equilibrium
configurations for the same $N$ and $\kappa$.

In actual computations, we have used $(r \times \theta \times \varphi)
= (128 \times 49 \times 81)$ grid points and we have taken into account the 
summation in the expansion of the Green function (\ref{green}) up to 24th
term, i.e. $\displaystyle P_{24}^{24}(\cos\theta) \cos 24\varphi$.
Although the integral region for integral equations (\ref{int rho}) -- 
(\ref{int omega}) extends to infinity, we have covered only a finite region 
($0 \le \tilde{r} \le 2$, $0 \le \theta \le \pi/2$, 
$0 \le \varphi \le \pi/2$) as was done for axisymmetric configurations 
in \cite{KEH89}.  In Newtonian gravity ($\kappa \sim 10^{-4}$), convergence 
is reached after 50 or so iterations, while in relativistic models 
($\kappa \sim 0.5$), it takes 200 or more iterations until converged 
solutions are obtained.

\subsection{Physical quantities}
Since there exist no exact equilibrium or stationary 
states, it is impossible to define conserved quantities for nonaxisymmetric
configurations in general relativity.  However, within our assumption of 
quasiequilibrium states or no gravitational waves, we can devise to define 
the approximate gravitational mass and the approximate angular momentum 
of the system.  

As mentioned before, since we neglect gravitational waves at infinity, 
the metric at infinity can be approximated by the flat spacetime.  Thus we 
assume that the metric function behaves as Equations (\ref{rho inf})--
(\ref{omega inf}).  Definitions of approximate quantities can be obtained 
as follows by taking the asymptotic behavior into consideration.

For the approximate angular momentum, from the field equation of $\omega$,
we can derive the following equation by arranging terms:
\begin{eqnarray*}
\lefteqn{\frac{1}{r^2}\pdif{}{r}\left(r^2\cdot r^2\sin^2\theta e^{3\beta-\nu}\pdif{\omega}{r}\right)%
+ \frac{1}{r\sin\theta}\pdif{}{\theta}\left(\sin\theta\cdot r^2\sin^2\theta e^{3\beta-\nu}\frac{1}{r}\pdif{\omega}{\theta}\right)}%
\hspace{2cm} \nonumber\\ &=&
-16\pi r\sin\theta e^{2\alpha+2\beta}\frac{(\varepsilon+p)v}{1-v^2}\nonumber\\
& &{}+2e^{2\alpha+\beta-\nu}\omega\left[2\ppdif{\alpha}{\varphi}+2(\pdif{\alpha}{\varphi})^2-2\pdif{\alpha}{\varphi}\left(\pdif{\beta}{\varphi}+\pdif{\nu}{\varphi}\right)\right] \ .
\end{eqnarray*}
By operating $\displaystyle \int r^2\sin\theta drd\theta d\varphi$ to this
equation, terms on the left hand side of this equation are converted into 
surface integrals and can be integrated as
\begin{eqnarray*}
\lefteqn{\int\left\{\pdif{}{r}\left(r^2\cdot r^2\sin^2\theta e^{3\beta-\nu}\pdif{\omega}{r}\right)dr\right\}\sin\theta d\theta d\varphi%
+\int\left\{\pdif{}{\theta}\left(\sin\theta\cdot r^2\sin^2\theta e^{3\beta-\nu}\frac{1}{r}\pdif{\omega}{\theta}\right)d\theta\right\}rdrd\varphi}%
\hspace{2cm}\nonumber\\ &=&
\int\left[r^4\sin^2\theta e^{3\beta-\nu}\pdif{\omega}{r}\right]_0^{\infty}
\sin\theta d\theta d\varphi%
+\int\left[\sin^3\theta e^{3\beta-\nu}\pdif{\omega}{\theta}\right]_0^{\pi} rdr d\varphi\\
&=&
\int\left(r^4e^{3\beta-\nu}\pdif{\omega}{r}\right)_{r=\infty}
\sin^3\theta d\theta d\varphi\\
&=&
-6J\int\sin^3\theta d\theta d\varphi\\
&=&
-16\pi J \ ,
\end{eqnarray*}
where we have used the asymptotic behavior $\displaystyle\omega \sim 
\frac{2J}{r^3}$ and $\displaystyle \pdif{\omega}{r}\sim -\frac{6J}{r^4}$ at 
$r\sim\infty$. By using this relation, we can find
\begin{eqnarray}
\label{angular momentum}
J &=& \int r\sin\theta e^{2\alpha+2\beta}\frac{(\varepsilon+p)v}{1-v^2} r^2\sin\theta dr d\theta d\varphi\nonumber\\
& &{}
-\frac{1}{4\pi}\int e^{2\alpha+\beta-\nu}\omega\left[\ppdif{\alpha}{\varphi}
+\pdif{\alpha}{\varphi}\left(\pdif{\alpha}{\varphi}-2\pdif{\beta}{\varphi}-2\pdif{\nu}{\varphi}\right)\right]r^2\sin\theta dr d\theta d\varphi \ .
\end{eqnarray}
Here we should note that in the asymptotic flat region, the coefficient of 
the first term of the expansion of the metric component $\omega$ 
in terms of $1/r$ is interpreted as the twice of the total 
angular momentum of the system.  Therefore, we may define the quantity $J$ as 
the approximate angular momentum of the nonaxisymmetric configuration.

In the same way, we can define the approximate gravitational mass as follows:
\begin{eqnarray}
M &=& \int e^{2\alpha+\beta+\nu}\left\{(\varepsilon +p)\frac{1+v^2}{1-v^2}+2p\right\} r^2\sin\theta dr d\theta d\varphi\nonumber\\
& &{}
+\int 2r\sin\theta e^{2\alpha+2\beta}(\varepsilon +p)\frac{v\omega}{1-v^2}r^2\sin\theta dr d\theta d\varphi \ ,
\end{eqnarray}
where we have used the asymptotic behavior $\displaystyle\omega \sim 
\frac{2J}{r^3}$ and $\displaystyle \pdif{\nu}{r}\sim\frac{M}{r^2}$ at 
$r \sim \infty$.

As for the rest mass, it is natural to define as:
\begin{eqnarray}
M_0 &=& \int \rho_0 u^t \sqrt{-g} \cdot r^2\sin\theta dr d\theta d\varphi\nonumber\\
&=& \int e^{2\alpha+\beta}\frac{\varepsilon}{(1+p/\varepsilon)^N}\frac{1}{\sqrt{1-v^2}} r^2\sin\theta dr d\theta d\varphi \ ,
\end{eqnarray}
where $\displaystyle \rho_0 = \frac{\varepsilon}{(1+p/\varepsilon)^N}$ is the 
baryon mass density.

\section{Results}
\subsection{Numerical tests of our new code}
We have checked our code by applying it to two cases for which solutions
have been obtained by other methods: 1) rotating and {\it general relativistic 
axisymmetric sequences} of polytropes and 2) {\it Newtonian binary sequences} 
of polytropes.

As mentioned before, our formulation becomes exact for axisymmetric 
configurations because the metric form (\ref{metric}) is the general one for 
the axisymmetric space-time and there can be exact stationary states.  
Komatsu et al.\cite{KEH89} computed general relativistic and axisymmetric 
rotating polytropes by developing a 2D code. We have computed the same 
equilibrium sequences by applying our 3D code.

In Figures \ref{axisymmetric} (a) and (b), we show equilibrium
sequences of models starting from $r_p/r_e=0.9375$ (nearly spherical) to 
$r_p/r_e=0.5625$ with $\kappa = 0.0001$, $0.25$ and $0.4$ for (a) $N=0$ 
and (b) $N = 0.5$, respectively, where $r_p$ and $r_e$ are the polar radius 
and the equatorial radius of axisymmetric configurations, respectively.
In these figures,  the squared nondimensional angular velocity 
$\tilde{\Omega}^2$ is plotted against the nondimensional angular momentum
$\tilde{J}$, where
\begin{equation}
\tilde{J} \equiv \frac{J}{(4\pi Gc^{2/3}M_0^{10/3}/\varepsilon_c^{1/3})^{1/2}}
\ .
\end{equation}
Three curves show our 3D computational results and discrete points denote the 
results of the 2D code.  As seen from these figures, our results of 3D code 
agree well with those of Komatsu et al.\cite{KEH89} to within less than 
$0.5 \%$.

As for binary sequences, we have carried out computations of Newtonian binary 
sequences. In our 3D code, Newtonian limit is treated by choosing 
$\kappa = 0.0001$. We have computed equilibrium sequences for
$N=0.0$, $0.5$, $1.0$ and $1.5$ from $q = 0.0$ (contact phase) to $q = 0.5$.
Our results are compared with those of Hachisu\cite{H86} in Figure 
\ref{Newton binary}.

As seen from this figure, except for $N=0.0$ models, corresponding models 
of two different codes agree well each other to within less than 
$0.5 \%$.
For the $N=0.0$ sequence, some of our models are not in good agreement with 
those of Newtonian computations.  This may be because our 3D code does not 
treat the surface region of the star carefully for the constant density 
distributions of $N = 0$ polytropes due to small mesh numbers within the 
star.  In our computations, the star is covered with 64 grid points at most 
in the $r-$direction and its number may not be enough to treat drastically 
changing density distributions. Although it is possible to increase grid 
points enough to manage a stiff distribution of the density, we did not try it 
because such a stiff equation of state is not suitable for real neutron stars.

\subsection{Synchronously rotating relativistic binary sequences}
In real evolution of compact binary systems, the matter cannot be represented 
by a simple polytrope and structures of compact stars will change during 
evolution so that the value of $\kappa$ must be changing. Such realistic 
models can be treated by using the realistic equation of state and following 
quasiequilibrium sequences with constant baryon mass models. 
However, in this paper, we report only the results of binary sequences with 
fixed values of $N$ and $\kappa$ because the main purpose of this paper 
is to show a new numerical scheme to treat nonaxisymmetric and
general relativistic configurations in quasiequilibrium states.

In Figures \ref{relativistic binary} (a), (b) and (c), we show the 
results of synchronously rotating binary equilibrium sequences 
from $q=0.0$ to $q = 0.5$ with $\kappa=0.0001$, $0.1$, $0.3$ and $0.5$, 
for (a) $N=0.0$, (b) $N=0.5$ and (c) $N=1.0$ polytropes, respectively,
(see also Tables in Appendix).

From these figures, we can see that as the strength of gravity is increased,
sequences are shifted to the part with a larger value of the dimensionless 
angular momentum, i.e. at the righter parts of the panels. Similar tendency
can be found for Newtonian binary sequences such as Figure \ref{Newton 
binary}. In the Newtonian models, models with larger values of $N$
are shifted to the righter part of the panel because of the density 
concentration to the central part of the star. As discussed in \cite{KEH89},
the effect of general relativity is to increase the mass concentration to the
central part of the star. In other words, models with large $\kappa$ can be
regarded as models with higher "effective" polytropic indices. Consequently
sequences of larger $\kappa$ models are located at the part with a larger 
value of the dimensionless angular momentum in the figure.  This situation can 
be clearly seen in Figure \ref{contours}. In this figure, equi-density 
contours are shown for $N = 0.5$ polytropes with $\kappa = 0.0001$ 
(Newtonian model) and $\kappa = 0.51$ (relativistic model).
For the same value of $N$, the density concentrates to the central region
of the stars for relativistic models more than that for Newtonian models.

The other characteristic feature of these figures is that for $N = 0$
sequences there are turning points where the value of the dimensionless
angular momentum becomes minimum but that for $N = 0.5$ and $N = 1.0$ sequences
there are no turning points except for Newtonian case with $N = 0.5$.

As discussed in Introduction, in the framework of Newtonian gravity, 
many sequences have been investigated. In particular, binary sequences are 
known to be connected to axisymmetric ones by way of Jacobi--Dumb-bell shaped 
sequences\cite{EHS82}. We show the whole relation of several equilibrium 
sequences of $N = 0$ polytropes, i.e. from Newtonian to relativistic and
from axisymmetric to binary sequences as well as the Jacobi--Dumb-bell shaped
sequence for Newtonian gravity in Figure \ref{overview}.

\section{Discussion and conclusions}
\subsection{Discussion}

As discussed before, the metric form (\ref{metric}) is too simplified 
and may not be suitable for 3D quasiequilibrium configurations. 
In general, the number of independent metric functions for stationary 
spacetime can be reduced to six. Thus, the most crucial point may be that we 
have not included nondiagonal components except for the $t \varphi$-component.
The results obtained in this paper may be affected by taking into account 
such nondiagonal components as well as letting $\alpha^{'} \ne \alpha$. 
It is not easy to estimate the effect of these
terms.  For stationary configurations, the Einstein equations for the 
nondiagonal metric components are written as follows (see e.g. \cite{LL}):
\begin{equation}
\pdif{(\sqrt{\sigma} h^{3/2} w^{ij})}{x^j} = 16 \pi h \sqrt{\sigma}
\frac{\varepsilon + p}{1 - v^2} v^i , \
\end{equation}
where
\begin{eqnarray}
 \sigma      & \equiv & {\rm det}  \ \sigma_{ij} \ , \\
 \sigma_{ij} & \equiv & g_{ij} + h g_i g_j \ , \\
 g_i         & \equiv & - \frac{g_{ti}}{g_{tt}} \ , \\
  h          & \equiv & - g_{tt} \ , \\
 w^{ij}      & \equiv & \sigma^{ik} \sigma^{jl} w_{kl} \ , \\
 w_{kl}      & \equiv & \pdif{g_k}{x^l} - \pdif{g_l}{x^k} \ , \\
 v^i         & \equiv & \frac{u^i}{\sqrt{h} (1- g_j u^j)} \ , \\
 v^2         & \equiv & \sigma_{ij} v^i v^j \ . 
\end{eqnarray}
These equations can be considered to be {\it linear equations}
for $g_i$ with source terms proportional to the velocity $v^i$ if $h$ and
$\sigma_{ij}$ are assumed to be known. This implies that even when
matter velocity is along the azimuthal direction, i.e. $v^{\varphi} \ne 0,
v^r = v^{\theta} = 0$, other nondiagonal components
of the metric will emerge.  Thus, in general, three nondiagonal
components may be of the same order because they are proportional
to the value of the 3-velocity.  However, since effect of the nondiagonal 
components is physically interpreted as dragging of the inertial
frame to the corresponding direction, it may be natural to consider that the 
nondiagonal component along the direction of the matter velocity dominates 
over nondiagonal components along two other directions.  Quantitative
estimation should be done by extending the present scheme or devising
new schemes.

Concerning the form of the metric, we need to discuss the relation
between our metric and the CFC scheme.
In the investigations which employ the CFC, the following form of the 
metric has been used\cite{WMM96,BCSST98a}:
\begin{equation}
ds^2 = -\alpha^2 dt^2 + \gamma_{ij} (dx^i -\omega^i dt)(dx^j -\omega^j dt) \ ,
\end{equation}
where 
\begin{equation}
\gamma_{ij} = \Psi^4 f_{ij} \ ,
\end{equation}
and Latin indices run from 1 to 3. Here $f_{ij}$ is the flat space metric and 
$\Psi$ is the conformal factor, which is independent of $t$. The spatial 
part of this metric is written as:
\begin{equation}
g_{ij}dx^idx^j = \Psi^4 (dr^2 + r^2 d\theta^2 + r^2 \sin^2\theta d\varphi^2) \ .
\end{equation}
On the other hand, the spatial part of our metric is:
\begin{equation}
g_{ij}dx^idx^j = e^{2\alpha}dr^2 + e^{2\alpha} r^2d\theta^2 + e^{2\beta} r^2\sin^2\theta d\varphi^2 \ .
\end{equation}
Therefore by comparing $\alpha$ and $\beta$ obtained from our code, we can 
see how the CFC is satisfied. Cook et al.\cite{CST96} did the same estimate
for axisymmetric models and found that the CFC is well satisfied for rapidly
rotating and general relativistic stars. 

In Figure \ref{metric distribution} distributions of the metric functions
in the meridional plane for a selected model are shown. In Figure 
\ref{difference} relative difference of
$|\exp(2\alpha) - \exp(2\beta)|/\exp(2\alpha)$ is plotted 
against the distance from the rotation axis. As seen from this figure, 
distributions of two metric components $\alpha$ and $\beta$ are not similar.
As far as $\kappa \leq 0.3$, relative differences between $\alpha$ 
and $\beta$ are within $5 \%$ not only for axisymmetric stars but also
for binary star systems.  Thus in this range of gravity strength, the CFC
may be a good approximation.  However, for models with $\kappa \sim 0.5$, 
deviations are larger than $10 \%$. We note that this conclusion is
obtained for our metric (\ref{metric}) which is not the most general one.
However, our results show that one should be careful when the CFC 
is used for extremely relativistic models.

The local flatness near the rotational axis requires that
\begin{equation}
\alpha = \beta \ , \ \ \ {\rm on \ \ the \ \ rotation \ \ axis}.
\end{equation}
From Figures \ref{metric distribution} and \ref{difference} the value of
$\alpha$ coincides with that of $\beta$ to within less than $2 \%$ 
on the rotational axis. Thus our numerical code satisfies the local flatness 
condition to this accuracy.

As seen from Equations (\ref{int rho})-(\ref{int omega}), we must cover the 
whole space from the origin to infinity in the integrations. In actual 
computations, however, we have covered only a finite region, i.e. the 
region of integration is limited to $0 \le \tilde{r} \le 2$. 
To check the error due to this truncation of the region, we have solved a 
few models by setting the size of the region to be twice as large as
the former cases : $0 \le \tilde{r} \le 4$. The difference of nondimensional 
physical quantities, such as the mass, the angular momentum and the angular 
velocity, between these two cases are less than $0.5 \%$. 
Concerning the approximate angular momentum, since its value depends on 
the distributions of the metric functions as seen from Equation 
(\ref{angular momentum}), the obtained approximate angular momentum contains 
some other error. When we change the integral region  from $0 \le \tilde{r} 
\le 2$ to $0 \le \tilde{r} \le 1$ in (\ref{angular momentum}), difference is 
less than $0.2 \%$. Our obtained values can be considered to be accurate 
enough.

As for the equation of state, we have chosen the relation (\ref{polytrope})
as our polytropes. Quantitative values in this paper are different from
those obtained from other choices of polytrope such as:
\begin{equation}
\varepsilon = \rho_0 + N p, \ \ \ \ \ \  p = K^{'} \rho_0^{1+1/N} \ .
\end{equation}

As shown in the previous section, although there exists a turning point
along the quasiequilibrium sequence for stiff equations of state,
no turning points appear for softer equations of state.  The existence
of turning points along {\it properly chosen} equilibrium sequences
are deeply related to the occurrence of some kind of instability (see 
e.g. \cite{UE98a,BCSST98b}). Therefore, it is important to develop numerical 
schemes which can compute unstable equilibrium configurations.  
In this sense, our present scheme is the one which is extended to 
handle realistic equilibrium sequences.

\subsection{Conclusion}

In this paper we have presented a new numerical scheme to handle 3D
configurations in quasiequilibrium states in general relativity.
By using the new scheme, we have succeeded in obtaining quasiequilibrium
sequences of synchronously rotating binary star systems.  We have treated 
restricted situations about 1) the metric, 2) the equation of state, and 3) 
the velocity field for the binary systems. The next step of our investigations 
is to remove some or all of these restrictions by extending the formulation to 
the more general form of the metric, including realistic equations of state 
and/or treating irrotational binary star systems. At that stage, we will be 
able to study turning points or critical configurations by applying the 
extended formulation.

\acknowledgments

We would like to thank Drs. Shogo Nishida, Shijun Yoshida and Shin'ichirou 
Yoshida for discussions. One of us (KU) would like to express his sincere 
gratitude to Professors Dennis W. Sciama and John C. Miller for their warm 
hospitality at SISSA.

\appendix

\section{Ricci tensor and Einstein equations in the tetrad system}

\def\pdif#1#2{\frac{\partial #1}{\partial #2}}
\def\dif#1#2{\frac{d #1}{d #2}}
\def\ppdif#1#2{\frac{\partial^2 #1}{\partial #2^2}}
\def\ddif#1#2{\frac{d^2 #1}{d #2^2}}

\def\gam{\gamma}
\def\alp{\alpha}
\def\omg{\omega}

\def\drhop{\pdif{\rho}{\varphi}}
\def\dgamp{\pdif{\gam}{\varphi}}
\def\dalpp{\pdif{\alp}{\varphi}}
\def\domgp{\pdif{\omg}{\varphi}}
\def\dbetap{\pdif{\beta}{\varphi}}
\def\dnup{\pdif{\nu}{\varphi}}

\def\drhot{\pdif{\rho}{\theta}}
\def\dgamt{\pdif{\gam}{\theta}}
\def\dalpt{\pdif{\alp}{\theta}}
\def\domgt{\pdif{\omg}{\theta}}
\def\dbetat{\pdif{\beta}{\theta}}
\def\dnut{\pdif{\nu}{\theta}}

\def\drhor{\pdif{\rho}{r}}
\def\dgamr{\pdif{\gam}{r}}
\def\dalpr{\pdif{\alp}{r}}
\def\domgr{\pdif{\omg}{r}}
\def\dbetar{\pdif{\beta}{r}}
\def\dnur{\pdif{\nu}{r}}

\def\ddrhop{\ppdif{\rho}{\varphi}}
\def\ddgamp{\ppdif{\gam}{\varphi}}
\def\ddalpp{\ppdif{\alp}{\varphi}}
\def\ddomgp{\ppdif{\omg}{\varphi}}
\def\ddbetap{\ppdif{\beta}{\varphi}}
\def\ddnup{\ppdif{\nu}{\varphi}}

\def\ddrhot{\ppdif{\rho}{\theta}}
\def\ddgamt{\ppdif{\gam}{\theta}}
\def\ddalpt{\ppdif{\alp}{\theta}}
\def\ddomgt{\ppdif{\omg}{\theta}}
\def\ddbetat{\ppdif{\beta}{\theta}}
\def\ddnut{\ppdif{\nu}{\theta}}

\def\ddrhor{\ppdif{\rho}{r}}
\def\ddgamr{\ppdif{\gam}{r}}
\def\ddalpr{\ppdif{\alp}{r}}
\def\ddomgr{\ppdif{\omg}{r}}
\def\ddbetar{\ppdif{\beta}{r}}
\def\ddnur{\ppdif{\nu}{r}}

\def\ddnurp{\frac{\partial^2 \nu}{\partial r \partial \varphi}}
\def\ddalprp{\frac{\partial^2 \alpha}{\partial r \partial \varphi}}
\def\ddbetarp{\frac{\partial^2 \beta}{\partial r \partial \varphi}}
\def\ddomegarp{\frac{\partial^2 \omega}{\partial r \partial \varphi}}

\def\ddnurt{\frac{\partial^2 \nu}{\partial r \partial \theta}}
\def\ddbetart{\frac{\partial^2 \beta}{\partial r \partial \theta}}

\def\ddnutp{\frac{\partial^2 \nu}{\partial \theta \partial \varphi}}
\def\ddalptp{\frac{\partial^2 \alpha}{\partial \theta \partial \varphi}}
\def\ddbetatp{\frac{\partial^2 \beta}{\partial \theta \partial \varphi}}
\def\ddomegatp{\frac{\partial^2 \omega}{\partial \theta \partial \varphi}}

\def\rrss{r^2\sin^2\theta}


Ricci tensors associated with the metric (\ref{metric}), $R_{(\zeta)(\eta)}$, 
is explicitly written in the tetrad system as follows:

\begin{eqnarray}
R_{(t)(t)} &=&%
e^{-2\alp}\left[\ddnur+\dnur\left(\dnur+\dbetar\right) + \frac{2}{r}\dnur\right]\nonumber\\
& &{}\hspace{1cm}%
+\frac{e^{-2\alpha}}{r^2}\left[\ddnut+\dnut\left(\dnut+\dbetat\right)+\cot\theta\dnut\right]\nonumber\\
& &{}\hspace{1cm}%
+\frac{e^{-2\beta}}{\rrss}\left[\ddnup+\dnup\left(\dnup-\dbetap\right)+2\dnup\dalpp\right]\nonumber\\
& &{}\hspace{1cm}%
-\frac{1}{2}e^{2\beta-2\nu-2\alp}\rrss \left[(\domgr)^2+\frac{1}{r^2}(\domgt)^2\right]\nonumber\\
& &{}\hspace{1cm}%
-e^{-2\nu}(\omg-\Omega)^2\left[2\ddalpp+\ddbetap+2(\dalpp)^2+\dbetap\left(\dbetap-\dnup\right)-2\dnup\dalpp\right]\nonumber\\
& &{}\hspace{1cm}%
-e^{-2\nu}(\omg-\Omega)\left[\ddomgp+2\domgp\dalpp+\domgp\left(3\dbetap-\dnup\right)\right]\nonumber\\
& &{}\hspace{1cm}%
-e^{-2\nu}(\domgp)^2 \ ,
\end{eqnarray}\\

\begin{eqnarray}
R_{(t)(r)} &=&%
-e^{-\alp-\nu}\left[(\omg-\Omega)\left(\ddalprp+\ddbetarp\right)+\frac{1}{2}\ddomegarp+\frac{1}{2}\domgr\left(\dbetap+\dnup\right)\right.\nonumber\\
& &{}\hspace{1.5cm}%
+(\omg-\Omega)\left(\dbetar-\dnur\right)\dbetap-(\omg-\Omega)\dalpp\left(\dnur+\dbetar\right)\nonumber\\
& &{}\hspace{1.5cm}%
\left.+\left(\dbetar-\dnur\right)\domgp+\frac{(\omg-\Omega)}{r}\left(\dbetap-\dalpp\right)+\frac{1}{r}\domgp\right] \ ,
\end{eqnarray}\\

\begin{eqnarray}
R_{(t)(\theta)} &=&%
-\frac{e^{-\alp-\nu}}{r}\left[(\omg-\Omega)\left(\ddalptp+\ddbetatp\right)+\frac{1}{2}\ddomegatp+\frac{1}{2}\domgt\left(\dbetap+\dnup\right)\right.\nonumber\\
& &{}\hspace{1cm}%
+(\omg-\Omega)\left(\dbetat-\dnut\right)\dbetap-(\omg-\Omega)\dalpp\left(\dnut+\dbetat\right)\nonumber\\
& &{}\hspace{1cm}%
\left.+\left(\dbetat-\dnut\right)\domgp+\cot\theta(\omg-\Omega)\left(\dalpp+\dbetap\right)+\cot\theta\domgp\right] \ ,
\end{eqnarray}\\

\begin{eqnarray}
R_{(t)(\varphi)} &=&%
\frac{1}{2}e^{\beta-\nu}r\sin\theta\left[e^{-2\alp}\ddomgr+e^{-2\alp}\left(3\dbetar+\frac{4}{r}-\dnur\right)\domgr\right.\nonumber\\
& &{}\hspace{2cm}%
+\left.\frac{e^{-2\alp}}{r^2}\left\{\ddomgt+\left(3\dbetat+3\cot\theta-\dnut\right)\domgt\right\}\right]\nonumber\\
& &{}\hspace{0cm}%
+\frac{e^{-\beta-\nu}}{r\sin\theta}(\omg-\Omega)\left[-2\ddalpp-2(\dalpp)^2+2\left(\dbetap+\dnup\right)\dalpp\right] \ ,
\end{eqnarray}\\

\begin{eqnarray}
R_{(r)(r)} &=&%
-e^{-2\alp}\left[\ddnur+\ddbetar+\ddalpr+(\dnur)^2+(\dbetar)^2-\left(\dnur+\dbetar\right)\dalpr+\frac{2}{r}\dbetar\right]\nonumber\\
& &{}\hspace{0cm}%
-\frac{e^{-2\alp}}{r^2}\left[\ddalpt+\cot\theta\dalpt+\dalpt\left(\dnut+\dbetat\right)\right]\nonumber\\
& &{}\hspace{0cm}%
-\frac{e^{-2\beta}}{\rrss}\left[\ddalpp+\left(\dnup-\dbetap\right)\dalpp+2(\dalpp)^2\right]\nonumber\\
& &{}\hspace{0cm}%
+e^{-2\nu}(\omg-\Omega)^2\left[\ddalpp-\left(\dnup-\dbetap\right)\dalpp+2(\dalpp)^2\right]\nonumber\\
& &{}\hspace{0cm}%
+2e^{-2\nu}(\omg-\Omega)\dalpp\domgp+\frac{1}{2}e^{2\beta-2\nu-2\alp}\rrss(\domgr)^2 \ ,
\end{eqnarray}\\

\begin{eqnarray}
R_{(r)(\theta)} &=&%
-\frac{e^{-2\alp}}{r}\left[\ddnurt + \ddbetart + \dnur\dnut+\dbetar\dbetat-(\dnur+\dbetar)\dalpt-(\dnut+\dbetat)\dalpr\right.\nonumber\\
& &{}\hspace{1.5cm}%
+\left.\cot\theta\dbetar-\cot\theta\dalpr-\frac{1}{r}\dnut-\frac{1}{r}\dalpt\right] \ ,
\end{eqnarray}\\

\begin{eqnarray}
R_{(r)(\varphi)} &=&%
-\frac{e^{-\alp-\beta}}{r\sin\theta}\left[\ddnurp +\ddalprp+\left(\dnur-\dbetar\right)\dnup-\left(\dnur+\dbetar\right)\dalpp-\frac{1}{r}\left(\dalpp+\dnup\right)\right.\nonumber\\
& &{}\hspace{1.5cm}%
-\frac{1}{2}e^{2\beta-2\nu}\rrss\left\{(\omg-\Omega)\ddomegarp+(\omg-\Omega)\domgr\left(3\dbetap-\dnup\right)+\left.2\domgr\domgp\right\}\right] \ ,\nonumber\\
\end{eqnarray}\\

\begin{eqnarray}
R_{(\theta)(\theta)}&=&%
-e^{-2\alp}\left[\ddalpr+\frac{2}{r}\dalpr+\left(\dalpr+\frac{1}{r}\right)\left(\dnur+\dbetar\right)\right.\nonumber\\
& &{}\hspace{0cm}%
-\frac{e^{-2\alp}}{r^2}\left[\ddalpt+\ddbetat+\ddnut+(\dnut)^2+(\dbetat)^2\right.\nonumber\\
& &{}\hspace{2cm}%
-\left.\dalpt\left(\dnut+\dbetat\right)-\cot\theta\dalpt+2\cot\theta\dbetat\right]\nonumber\\
& &{}\hspace{0cm}%
-\frac{e^{-2\beta}}{\rrss}\left[\ddalpp+2(\dalpp)^2+\dalpp\left(\dnup-\dbetap\right)\right]\nonumber\\
& &{}\hspace{0cm}%
+e^{-2\nu}(\omg-\Omega)^2\left[\ddalpp+2(\dalpp)^2-\dalpp\left(\dnup-\dbetap\right)\right]\nonumber\\
& &{}\hspace{0cm}%
+2e^{-2\nu}(\omg-\Omega)\dalpp\domgp+\frac{1}{2}e^{2\beta-2\nu-2\alp}\sin^2\theta(\domgt)^2 \ ,
\end{eqnarray}\\

\begin{eqnarray}
R_{(\theta)(\varphi)}&=&%
-\frac{e^{-\beta-\alp}}{\rrss}\left[\ddnutp + \ddalptp +\left(\dnut-\dbetat-\cot\theta\right)\dnup-\left(\dnut+\dbetat+\cot\theta\right)\dalpp\right.\nonumber\\
& &{}\hspace{1.5cm}%
-\frac{1}{2}e^{2\beta-2\nu}\rrss\left\{(\omg-\Omega)\ddomegatp+(\omg-\Omega)\domgt\left(3\dbetap-\dnup\right)+\left.2\domgt\domgp\right\}\right] \ ,\nonumber\\
\end{eqnarray}\\

\begin{eqnarray}
R_{(\varphi)(\varphi)} &=&%
-e^{-2\alp}\left[\ddbetar+\frac{3}{r}\dbetar+\dbetar\left(\dbetar+\dalpr\right)+\frac{1}{r}\dnur+\frac{1}{r^2}\right]\nonumber\\
& &{}\hspace{0cm}%
-\frac{e^{-2\alp}}{r^2}\left[\ddbetat+2\cot\theta\dbetat+\dbetat\left(\dbetat+\dnut\right)+\cot\theta\dnut-1\right]\nonumber\\
& &{}\hspace{0cm}%
-\frac{e^{-2\beta}}{\rrss}\left[2\ddalpp+\ddnup+2(\dalpp)^2-2\dbetap\dalpp+\dnup\left(\dnup-\dbetap\right)\right]\nonumber\\
& &{}\hspace{0cm}%
+e^{-2\nu}(\omg-\Omega)^2\left[\ddbetap+\dbetap\left(\dbetap-\dnup\right)+2\dbetap\dalpp\right]\nonumber\\
& &{}\hspace{0cm}%
+e^{-2\nu}(\omg-\Omega)\left[\ddomgp+\domgp\left(3\dbetap-\dnup+2\dalpp\right)\right]\nonumber\\
& &{}\hspace{0cm}%
+e^{-2\nu}(\domgp)^2-\frac{1}{2}e^{2\beta-2\nu-2\alp}\rrss\left\{(\domgr)^2+\frac{1}{r^2}(\domgt)^2\right\} \ ,
\end{eqnarray}

where we have used the orthonormal tetrad defined as:

\begin{equation}
\displaystyle\lambda_{(\eta)}^{\xi} = \left(\begin{array}{llll}
e^{-\nu},& 0,&  0,&  (\omega-\Omega) e^{-\nu}\\
0,& e^{-\alpha},& 0,& 0\\
0,& 0,& \frac{e^{-\alpha}}{r},& 0\\
0,& 0,& 0,& \frac{e^{-\beta}}{r\sin\theta}\\
\end{array}\right) \ ,
\end{equation}\\

and

\begin{equation}
\displaystyle\lambda_{(\eta)_\xi} = \left(\begin{array}{llll}
-e^{-\nu},& 0,& 0,& 0\\
0,& e^{\alpha},& 0,& 0\\
0,& 0,& re^{\alpha},& 0\\
-r\sin\theta e^{\beta}(\omg-\Omega),& 0,& 0,& r\sin\theta e^{\beta}\\
\end{array}\right) \ .
\end{equation}\\

Any tensor, $h_{\zeta \eta}$, is transformed to that in the tetrad system
as follows:
\begin{equation}
h_{(\alpha)(\beta)} = h_{\zeta \eta} \lambda_{(\alpha)}^{\zeta}\lambda_{(\beta)}^{\eta} \ .
\end{equation}


The energy-momentum tensor for the quasiequilibrium configurations in this 
paper are:

\begin{eqnarray}
T_{(t)(t)} &=& \frac{\varepsilon+pv^2}{1-v^2} \ ,\\
T_{(r)(r)} = T_{(\theta)(\theta)} &=& p \ , \\
T_{(\varphi)(\varphi)} &=& \frac{p+\varepsilon v^2}{1-v^2} \ , \\
T_{(t)(\varphi)} = T_{(\varphi)(t)} &=& -\frac{(\varepsilon+p)v}{1-v^2} \ , \\
\mbox{others} &=& 0 \ .
\end{eqnarray}

Einstein equation is formally written as:

\begin{equation}
R_{(\zeta)(\eta)} = 8\pi (T_{(\zeta)(\eta)} - \frac{1}{2} g_{(\zeta)(\eta)} T) \ ,
\end{equation}
where $T = T^{(\zeta)}_{(\zeta)}$. We have used the following combinations to 
get elliptic type equations (\ref{rho})--(\ref{alpha}) and (\ref{omega new}).

\begin{eqnarray}
R_{(t)(t)}+R_{(\varphi)(\varphi)} &=& 8\pi (\varepsilon+p)\frac{1+v^2}{1-v^2} \ ,\\
R_{(t)(t)}-R_{(\varphi)(\varphi)} &=& 16\pi p \ , \\
R_{(t)(t)}+R_{(r)(r)}+R_{(\theta)(\theta)}-R_{(\varphi)(\varphi)} &=& 8\pi (\varepsilon +p) \ , \\
R_{(t)(\varphi)} &=& 8\pi (\varepsilon +p)\frac{v}{1-v^2} \ ,
\end{eqnarray}

\section{Detailed numerical results for binary sequences}

Numerical results of binary sequences are shown in Tables
\ref{table n=0.0 k=0.0001}-\ref{table n=1.0 k=0.5}.

In these Tables, $\tilde{r}_A$, $\tilde{J}$, $\tilde{M}$, $\tilde{M}_0$, 
$\tilde{\Omega}^2$ and $f_x$ are the dimensionless distance to point A from 
the origin, the dimensionless total angular momentum, the dimensionless total 
gravitational mass, the dimensionless total rest mass, the dimensionless 
angular velocity and the scale parameter, respectively.
Here, masses are normalized as:
\begin{eqnarray}
{\tilde M} \equiv \frac{M}{f_x^{3/2} c^4 /(G^{3/2} \varepsilon_c^{1/2})}  \ , \\
{\tilde M}_0 \equiv \frac{M_0}{f_x^{3/2} c^4 /(G^{3/2} \varepsilon_c^{1/2})}  \ .  
\end{eqnarray}
%


%
%

\begin{figure}
\caption{Schematic view of the binary system. 
The $x-$axis is set along the line joining the two centers of mass of the two
stars and the origin is set on the middle point. The $z-$axis is along
the rotation axis. Point A, B are set at the intersections of the surface 
of the star and the $x-$axis. The inner intersection is point A and 
the outer one is point B.}
\label{coordinate system}
\end{figure}

\begin{figure}
\caption{Dimensionless squared angular velocity 
$\Omega^2/(4\pi G\varepsilon_c/c^2)$ is plotted against dimensionless angular 
momentum $J/(4\pi G c^{2/3}M_0^{10/3}/\varepsilon_c^{1/3})^{1/2}$ for
the sequences of the axisymmetric stars from $r_p/r_e=0.9375$ (nearly 
spherical) to $r_p/r_e=0.5625$ with polytropic indices (a) $N=0.0$ and 
(b) $N=0.5$, where $r_p$ and $r_e$ are the polar radius and the equatorial 
radius of axisymmetric configurations. The solid line ($\kappa = 0.0001$), 
the dotted line ($\kappa = 0.25$) and the dashed line ($\kappa = 0.4$) are 
results obtained by using our 3D code. Crosses ($\kappa = 0.0001$), open 
circles ($\kappa = 0.25$) and filled circles ($\kappa = 0.4$) are taken from 
Komatsu et al.\protect\cite{KEH89}.}
\label{axisymmetric}
\end{figure}

\begin{figure}
\caption{Same as Figure \ref{axisymmetric} but for the sequences of Newtonian
binary systems from $q =0.0$ to $q = 0.5$. The solid line ($N=0.0$), 
the dotted line ($N=0.5$), the short-dashed line ($N=1.0$) and the long-dashed
line ($N=1.5$) are our numerical results. Crosses ($N=0.0$), filled circles 
($N=0.5$), open circles ($N=1.0$) and filled rectangles ($N=1.5$) are taken 
from Hachisu\protect\cite{H86}.}
\label{Newton binary}
\end{figure}

\begin{figure}	
\caption{Same as Figure \ref{axisymmetric} but for the sequences of 
relativistic binary systems from $q = 0.0$ to $q = 0.5$ with polytropic 
indices (a) $N=0.0$, (b) $N=0.5$ and (c) $N=1.0$. On each panel, 
$\kappa = 0.0001$ (Newtonian), $0.1$, $0.3$ 
and $0.5$ are shown by solid lines. Crosses denote the Newtonian results 
taken from Hachisu\protect\cite{H86}.}
\label{relativistic binary}
\end{figure}

\begin{figure}
\caption{Contours of the energy density on the $x-z$ plane.
The units of the distance is $r_B$.
(a) The contact phase ($q = 0.0$) and (b) the distant phase ($q=0.5$) are
shown for $N=0.5$ polytrope. On each panel, contours for $\kappa=0.0001$ 
(Newtonian) and $\kappa=0.5$ (relativistic) are shown.}
\label{contours}
\end{figure}

\begin{figure}
\caption{Same as Figure \ref{axisymmetric} but for $N = 0.0$
polytropes with different $\kappa$ and different topology, i.e. axisymmetric
sequences, Jacobi--Dumb-bell shaped sequence and binary sequences.
Solid, dotted and dashed curves denote sequences with $\kappa=0.0001$ 
(Newtonian), $0.25$ and $0.4$, respectively. The dash-dotted curve denotes 
the Jacobi--Dumb-bell shaped sequences taken from Hachisu\protect\cite{H86}.}
\label{overview}
\end{figure}

\begin{figure}
\caption{(a) Distribution of the metric function $\exp(2\nu)$ on the
equatorial plane is plotted against the distance from the rotational axis.
Model parameters are $N = 0.5$, $\kappa = 0.5$ and $q = 0.0$. Different curves 
correspond to distributions on $\varphi = 0, \pi/12, 2\pi/12, 3\pi/12, 
4\pi/12, 5\pi/12, 6\pi/12$.
(b) Distribution of the metric function $\exp(2\beta)$.
(c) Distribution of the metric function $\exp(2\alpha)$.
(d) Distribution of the metric function $\omega/\Omega$.
}
\label{metric distribution}
\end{figure}

\begin{figure}
\caption{Distribution of $|\exp(2\alpha) - \exp(2\beta)|/\exp(2\alpha)$
on the equatorial plane is plotted against the distance from the rotational
axis. Model parameters are $N = 0.5$, $\kappa = 0.5$ and $q = 0.0$. Different
curves correspond to distributions on $\varphi = 0, 3\pi/12, 6\pi/12$. 
}
\label{difference}
\end{figure}

%
%

\begin{table}
\caption{$N=0.0$, $\kappa=0.0001$ (Newtonian)}
\label{table n=0.0 k=0.0001}
\begin{tabular}{llllll}
$\tilde{r}_A$ & $\tilde{J}$ & $\tilde{M}$ & $\tilde{M}_0$ & $\tilde{\Omega}^2$ & $f_x$\\\hline
1.000E-08 & 1.126E-01 & 3.246E-06 & 3.246E-06 & 2.647E-02 & 3.984E-04\\
3.125E-02 & 1.121E-01 & 3.233E-06 & 3.234E-06 & 2.644E-02 & 3.960E-04\\
6.250E-02 & 1.109E-01 & 3.187E-06 & 3.187E-06 & 2.586E-02 & 3.894E-04\\
9.375E-02 & 1.097E-01 & 3.137E-06 & 3.138E-06 & 2.465E-02 & 3.836E-04\\
1.250E-01 & 1.082E-01 & 3.078E-06 & 3.078E-06 & 2.275E-02 & 3.797E-04\\
1.562E-01 & 1.078E-01 & 3.032E-06 & 3.033E-06 & 2.077E-02 & 3.830E-04\\
1.875E-01 & 1.074E-01 & 2.991E-06 & 2.992E-06 & 1.849E-02 & 3.893E-04\\
2.188E-01 & 1.078E-01 & 2.965E-06 & 2.965E-06 & 1.631E-02 & 4.013E-04\\
2.500E-01 & 1.087E-01 & 2.938E-06 & 2.938E-06 & 1.423E-02 & 4.176E-04\\
2.812E-01 & 1.096E-01 & 2.905E-06 & 2.906E-06 & 1.213E-02 & 4.379E-04\\
3.125E-01 & 1.112E-01 & 2.874E-06 & 2.874E-06 & 1.033E-02 & 4.620E-04\\
3.438E-01 & 1.140E-01 & 2.825E-06 & 2.826E-06 & 8.790E-03 & 4.914E-04\\
3.750E-01 & 1.150E-01 & 2.780E-06 & 2.780E-06 & 7.110E-03 & 5.261E-04\\
4.062E-01 & 1.182E-01 & 2.752E-06 & 2.753E-06 & 5.880E-03 & 5.713E-04\\
4.375E-01 & 1.212E-01 & 2.750E-06 & 2.750E-06 & 4.808E-03 & 6.262E-04\\
4.688E-01 & 1.201E-01 & 2.780E-06 & 2.781E-06 & 3.535E-03 & 7.014E-04 \\
5.000E-01 & 1.295E-01 & 2.820E-06 & 2.821E-06 & 3.089E-03 & 7.916E-04\\
\end{tabular}
\end{table}

\begin{table}
\caption{$N=0.5$, $\kappa=0.0001$ (Newtonian)}
\label{table n=0.5 k=0.0001}
\begin{tabular}{llllll}
$\tilde{r}_A$ & $\tilde{J}$ & $\tilde{M}$ & $\tilde{M}_0$ & $\tilde{\Omega}^2$ & $f_x$\\\hline
1.000E-08 & 1.145E-01 & 4.342E-06 & 4.342E-06 & 1.668E-02 & 6.474E-04 \\
3.125E-02 & 1.142E-01 & 4.330E-06 & 4.330E-06 & 1.647E-02 & 6.464E-04 \\
6.250E-02 & 1.137E-01 & 4.297E-06 & 4.298E-06 & 1.582E-02 & 6.455E-04 \\
9.375E-02 & 1.132E-01 & 4.256E-06 & 4.257E-06 & 1.480E-02 & 6.477E-04 \\
1.250E-01 & 1.130E-01 & 4.209E-06 & 4.210E-06 & 1.353E-02 & 6.549E-04 \\
1.562E-01 & 1.131E-01 & 4.168E-06 & 4.169E-06 & 1.213E-02 & 6.684E-04 \\
1.875E-01 & 1.136E-01 & 4.133E-06 & 4.134E-06 & 1.070E-02 & 6.891E-04 \\
2.188E-01 & 1.145E-01 & 4.101E-06 & 4.102E-06 & 9.309E-03 & 7.168E-04 \\
2.500E-01 & 1.158E-01 & 4.071E-06 & 4.072E-06 & 7.998E-03 & 7.523E-04 \\
2.812E-01 & 1.175E-01 & 4.041E-06 & 4.042E-06 & 6.800E-03 & 7.958E-04 \\
3.125E-01 & 1.195E-01 & 4.009E-06 & 4.010E-06 & 5.711E-03 & 8.484E-04 \\
3.438E-01 & 1.219E-01 & 3.980E-06 & 3.980E-06 & 4.746E-03 & 9.116E-04 \\
3.750E-01 & 1.246E-01 & 3.957E-06 & 3.957E-06 & 3.900E-03 & 9.877E-04 \\
4.062E-01 & 1.278E-01 & 3.948E-06 & 3.949E-06 & 3.188E-03 & 1.079E-03 \\
4.375E-01 & 1.316E-01 & 3.955E-06 & 3.955E-06 & 2.592E-03 & 1.190E-03 \\
4.688E-01 & 1.349E-01 & 3.968E-06 & 3.969E-06 & 2.052E-03 & 1.325E-03 \\
5.000E-01 & 1.408E-01 & 3.982E-06 & 3.982E-06 & 1.666E-03 & 1.486E-03 \\
\end{tabular}
\end{table}

\begin{table}
\caption{$N=1.0$, $\kappa=0.0001$ (Newtonian)}
\label{table n=1.0 k=0.0001}
\begin{tabular}{llllll}
$\tilde{r}_A$ & $\tilde{J}$ & $\tilde{M}$ & $\tilde{M}_0$ & $\tilde{\Omega}^2$ & $f_x$\\\hline
1.000E-08 & 1.182E-01 & 5.336E-06 & 5.337E-06 & 9.957E-03 & 1.028E-03 \\
3.125E-02 & 1.182E-01 & 5.327E-06 & 5.328E-06 & 9.802E-03 & 1.030E-03 \\
6.250E-02 & 1.183E-01 & 5.303E-06 & 5.304E-06 & 9.359E-03 & 1.037E-03 \\
9.375E-02 & 1.184E-01 & 5.274E-06 & 5.274E-06 & 8.702E-03 & 1.051E-03 \\
1.250E-01 & 1.189E-01 & 5.238E-06 & 5.239E-06 & 7.903E-03 & 1.073E-03 \\
1.562E-01 & 1.197E-01 & 5.206E-06 & 5.207E-06 & 7.044E-03 & 1.105E-03 \\
1.875E-01 & 1.208E-01 & 5.178E-06 & 5.178E-06 & 6.181E-03 & 1.147E-03 \\
2.188E-01 & 1.223E-01 & 5.148E-06 & 5.149E-06 & 5.345E-03 & 1.201E-03 \\
2.500E-01 & 1.241E-01 & 5.121E-06 & 5.122E-06 & 4.568E-03 & 1.268E-03 \\
2.812E-01 & 1.264E-01 & 5.096E-06 & 5.097E-06 & 3.861E-03 & 1.349E-03 \\
3.125E-01 & 1.289E-01 & 5.073E-06 & 5.074E-06 & 3.231E-03 & 1.446E-03 \\
3.438E-01 & 1.319E-01 & 5.055E-06 & 5.056E-06 & 2.679E-03 & 1.561E-03 \\
3.750E-01 & 1.352E-01 & 5.043E-06 & 5.043E-06 & 2.201E-03 & 1.697E-03 \\
4.062E-01 & 1.390E-01 & 5.037E-06 & 5.038E-06 & 1.796E-03 & 1.859E-03 \\
4.375E-01 & 1.431E-01 & 5.035E-06 & 5.036E-06 & 1.451E-03 & 2.052E-03 \\
4.688E-01 & 1.477E-01 & 5.031E-06 & 5.032E-06 & 1.158E-03 & 2.281E-03 \\
5.000E-01 & 1.532E-01 & 5.021E-06 & 5.022E-06 & 9.183E-04 & 2.558E-03 \\
\end{tabular}
\end{table}

\begin{table}
\caption{$N=1.5$, $\kappa=0.0001$ (Newtonian)}
\label{table n=1.5 k=0.0001}
\begin{tabular}{llllll}
$\tilde{r}_A$ & $\tilde{J}$ & $\tilde{M}$ & $\tilde{M}_0$ & $\tilde{\Omega}^2$ & $f_x$\\\hline
1.000E-08 & 1.246E-01 & 6.282E-06 & 6.282E-06 & 5.596E-03 & 1.662E-03 \\
3.125E-02 & 1.247E-01 & 6.278E-06 & 6.278E-06 & 5.509E-03 & 1.667E-03 \\
6.250E-02 & 1.251E-01 & 6.260E-06 & 6.261E-06 & 5.252E-03 & 1.684E-03 \\
9.375E-02 & 1.258E-01 & 6.240E-06 & 6.240E-06 & 4.878E-03 & 1.714E-03 \\
1.250E-01 & 1.268E-01 & 6.214E-06 & 6.215E-06 & 4.422E-03 & 1.758E-03 \\
1.562E-01 & 1.281E-01 & 6.191E-06 & 6.192E-06 & 3.936E-03 & 1.818E-03 \\
1.875E-01 & 1.297E-01 & 6.170E-06 & 6.170E-06 & 3.448E-03 & 1.896E-03 \\
2.188E-01 & 1.318E-01 & 6.146E-06 & 6.146E-06 & 2.975E-03 & 1.993E-03 \\
2.500E-01 & 1.342E-01 & 6.124E-06 & 6.125E-06 & 2.536E-03 & 2.110E-03 \\
2.812E-01 & 1.370E-01 & 6.105E-06 & 6.106E-06 & 2.140E-03 & 2.252E-03 \\
3.125E-01 & 1.401E-01 & 6.089E-06 & 6.089E-06 & 1.788E-03 & 2.419E-03 \\
3.438E-01 & 1.437E-01 & 6.075E-06 & 6.076E-06 & 1.481E-03 & 2.617E-03 \\
3.750E-01 & 1.476E-01 & 6.064E-06 & 6.065E-06 & 1.215E-03 & 2.850E-03 \\
4.062E-01 & 1.519E-01 & 6.055E-06 & 6.056E-06 & 9.891E-04 & 3.125E-03 \\
4.375E-01 & 1.567E-01 & 6.046E-06 & 6.046E-06 & 7.965E-04 & 3.453E-03 \\
4.688E-01 & 1.621E-01 & 6.036E-06 & 6.036E-06 & 6.353E-04 & 3.844E-03 \\
5.000E-01 & 1.680E-01 & 6.024E-06 & 6.024E-06 & 5.006E-04 & 4.314E-03 \\
\end{tabular}
\end{table}

\begin{table}
\caption{$N=0.0$, $\kappa=0.1$}
\label{table n=0.0 k=0.1}
\begin{tabular}{llllll}
$\tilde{r}_A$ & $\tilde{J}$ & $\tilde{M}$ & $\tilde{M}_0$ & $\tilde{\Omega}^2$ & $f_x$\\\hline
1.000E-08 & 1.316E-01 & 5.411E-02 & 5.997E-02 & 2.794E-02 & 1.531E-01 \\
3.125E-02 & 1.307E-01 & 5.386E-02 & 5.968E-02 & 2.742E-02 & 1.533E-01 \\
6.250E-02 & 1.296E-01 & 5.371E-02 & 5.953E-02 & 2.716E-02 & 1.519E-01 \\
9.375E-02 & 1.286E-01 & 5.317E-02 & 5.892E-02 & 2.553E-02 & 1.528E-01 \\
1.250E-01 & 1.274E-01 & 5.263E-02 & 5.829E-02 & 2.375E-02 & 1.543E-01 \\
1.562E-01 & 1.268E-01 & 5.215E-02 & 5.774E-02 & 2.145E-02 & 1.579E-01 \\
1.875E-01 & 1.270E-01 & 5.177E-02 & 5.728E-02 & 1.912E-02 & 1.634E-01 \\
2.188E-01 & 1.280E-01 & 5.153E-02 & 5.697E-02 & 1.687E-02 & 1.712E-01 \\
2.500E-01 & 1.291E-01 & 5.132E-02 & 5.672E-02 & 1.463E-02 & 1.804E-01 \\
2.812E-01 & 1.306E-01 & 5.117E-02 & 5.653E-02 & 1.250E-02 & 1.922E-01 \\
3.125E-01 & 1.325E-01 & 5.089E-02 & 5.619E-02 & 1.056E-02 & 2.060E-01 \\
3.438E-01 & 1.351E-01 & 5.056E-02 & 5.578E-02 & 8.922E-03 & 2.218E-01 \\
3.750E-01 & 1.375E-01 & 5.012E-02 & 5.526E-02 & 7.359E-03 & 2.415E-01 \\
4.062E-01 & 1.405E-01 & 4.981E-02 & 5.489E-02 & 5.996E-03 & 2.653E-01 \\
4.375E-01 & 1.439E-01 & 4.991E-02 & 5.500E-02 & 4.883E-03 & 2.929E-01 \\
4.688E-01 & 1.435E-01 & 5.020E-02 & 5.543E-02 & 3.631E-03 & 3.290E-01 \\
5.000E-01 & 1.556E-01 & 5.070E-02 & 5.584E-02 & 3.180E-03 & 3.713E-01 \\
\end{tabular}
\end{table}

\begin{table}
\caption{$N=0.5$, $\kappa=0.1$}
\label{table n=0.5 k=0.1}
\begin{tabular}{llllll}
$\tilde{r}_A$ & $\tilde{J}$ & $\tilde{M}$ & $\tilde{M}_0$ & $\tilde{\Omega}^2$ & $f_x$\\\hline
1.000E-08 & 1.438E-01 & 7.006E-02 & 7.664E-02 & 1.631E-02 & 2.457E-01 \\
3.125E-02 & 1.436E-01 & 6.998E-02 & 7.655E-02 & 1.608E-02 & 2.464E-01 \\
6.250E-02 & 1.432E-01 & 6.975E-02 & 7.629E-02 & 1.547E-02 & 2.486E-01 \\
9.375E-02 & 1.431E-01 & 6.943E-02 & 7.593E-02 & 1.443E-02 & 2.533E-01 \\
1.250E-01 & 1.432E-01 & 6.912E-02 & 7.557E-02 & 1.318E-02 & 2.602E-01 \\
1.562E-01 & 1.437E-01 & 6.882E-02 & 7.523E-02 & 1.180E-02 & 2.699E-01 \\
1.875E-01 & 1.448E-01 & 6.859E-02 & 7.495E-02 & 1.040E-02 & 2.826E-01 \\
2.188E-01 & 1.462E-01 & 6.842E-02 & 7.475E-02 & 9.029E-03 & 2.983E-01 \\
2.500E-01 & 1.481E-01 & 6.831E-02 & 7.461E-02 & 7.750E-03 & 3.174E-01 \\
2.812E-01 & 1.505E-01 & 6.819E-02 & 7.446E-02 & 6.574E-03 & 3.404E-01 \\
3.125E-01 & 1.533E-01 & 6.805E-02 & 7.427E-02 & 5.518E-03 & 3.679E-01 \\
3.438E-01 & 1.564E-01 & 6.791E-02 & 7.411E-02 & 4.582E-03 & 4.003E-01 \\
3.750E-01 & 1.598E-01 & 6.783E-02 & 7.399E-02 & 3.760E-03 & 4.386E-01 \\
4.062E-01 & 1.640E-01 & 6.784E-02 & 7.399E-02 & 3.068E-03 & 4.840E-01 \\
4.375E-01 & 1.689E-01 & 6.799E-02 & 7.414E-02 & 2.487E-03 & 5.376E-01 \\
4.688E-01 & 1.731E-01 & 6.820E-02 & 7.439E-02 & 1.962E-03 & 6.022E-01 \\
5.000E-01 & 1.807E-01 & 6.842E-02 & 7.459E-02 & 1.588E-03 & 6.787E-01 \\
\end{tabular}
\end{table}

\begin{table}
\caption{$N=1.0$, $\kappa=0.1$}
\label{table n=1.0 k=0.1}
\begin{tabular}{llllll}
$\tilde{r}_A$ & $\tilde{J}$ & $\tilde{M}$ & $\tilde{M}_0$ & $\tilde{\Omega}^2$ & $f_x$\\\hline
1.000E-08 & 1.612E-01 & 8.428E-02 & 9.030E-02 & 8.716E-03 & 4.072E-01 \\
3.125E-02 & 1.631E-01 & 8.461E-02 & 9.060E-02 & 8.861E-03 & 4.080E-01 \\
6.250E-02 & 1.618E-01 & 8.435E-02 & 9.037E-02 & 8.276E-03 & 4.151E-01 \\
9.375E-02 & 1.624E-01 & 8.421E-02 & 9.022E-02 & 7.712E-03 & 4.252E-01 \\
1.250E-01 & 1.634E-01 & 8.407E-02 & 9.005E-02 & 7.022E-03 & 4.397E-01 \\
1.562E-01 & 1.649E-01 & 8.395E-02 & 8.992E-02 & 6.269E-03 & 4.588E-01 \\
1.875E-01 & 1.668E-01 & 8.384E-02 & 8.979E-02 & 5.503E-03 & 4.829E-01 \\
2.188E-01 & 1.691E-01 & 8.378E-02 & 8.971E-02 & 4.763E-03 & 5.122E-01 \\
2.500E-01 & 1.719E-01 & 8.375E-02 & 8.967E-02 & 4.072E-03 & 5.474E-01 \\
2.812E-01 & 1.751E-01 & 8.375E-02 & 8.966E-02 & 3.443E-03 & 5.891E-01 \\
3.125E-01 & 1.788E-01 & 8.375E-02 & 8.965E-02 & 2.879E-03 & 6.386E-01 \\
3.438E-01 & 1.830E-01 & 8.376E-02 & 8.965E-02 & 2.385E-03 & 6.969E-01 \\
3.750E-01 & 1.876E-01 & 8.381E-02 & 8.969E-02 & 1.957E-03 & 7.653E-01 \\
4.062E-01 & 1.929E-01 & 8.391E-02 & 8.979E-02 & 1.593E-03 & 8.456E-01 \\
4.375E-01 & 1.986E-01 & 8.404E-02 & 8.992E-02 & 1.284E-03 & 9.405E-01 \\
4.688E-01 & 2.048E-01 & 8.417E-02 & 9.007E-02 & 1.023E-03 & 1.054E+00 \\
5.000E-01 & 2.119E-01 & 8.429E-02 & 9.018E-02 & 8.073E-04 & 1.190E+00 \\
\end{tabular}
\end{table}

\begin{table}
\caption{$N=0.0$, $\kappa=0.3$}
\label{table n=0.0 k=0.3}
\begin{tabular}{llllll}
$\tilde{r}_A$ & $\tilde{J}$ & $\tilde{M}$ & $\tilde{M}_0$ & $\tilde{\Omega}^2$ & $f_x$\\\hline
1.000E-08 & 1.638E-01 & 1.231E-01 & 1.469E-01 & 2.870E-02 & 1.447E-01 \\
3.125E-02 & 1.629E-01 & 1.227E-01 & 1.466E-01 & 2.825E-02 & 1.452E-01 \\
6.250E-02 & 1.620E-01 & 1.223E-01 & 1.461E-01 & 2.773E-02 & 1.457E-01 \\
9.375E-02 & 1.613E-01 & 1.217E-01 & 1.455E-01 & 2.604E-02 & 1.491E-01 \\
1.250E-01 & 1.600E-01 & 1.212E-01 & 1.450E-01 & 2.406E-02 & 1.536E-01 \\
1.562E-01 & 1.598E-01 & 1.209E-01 & 1.448E-01 & 2.170E-02 & 1.600E-01 \\
1.875E-01 & 1.603E-01 & 1.206E-01 & 1.445E-01 & 1.919E-02 & 1.688E-01 \\
2.188E-01 & 1.609E-01 & 1.206E-01 & 1.448E-01 & 1.674E-02 & 1.790E-01 \\
2.500E-01 & 1.627E-01 & 1.208E-01 & 1.450E-01 & 1.441E-02 & 1.924E-01 \\
2.812E-01 & 1.651E-01 & 1.210E-01 & 1.453E-01 & 1.240E-02 & 2.073E-01 \\
3.125E-01 & 1.673E-01 & 1.210E-01 & 1.455E-01 & 1.040E-02 & 2.260E-01 \\
3.438E-01 & 1.697E-01 & 1.208E-01 & 1.453E-01 & 8.631E-03 & 2.479E-01 \\
3.750E-01 & 1.725E-01 & 1.206E-01 & 1.451E-01 & 7.139E-03 & 2.728E-01 \\
4.062E-01 & 1.755E-01 & 1.204E-01 & 1.450E-01 & 5.749E-03 & 3.048E-01 \\
4.375E-01 & 1.778E-01 & 1.205E-01 & 1.454E-01 & 4.562E-03 & 3.405E-01 \\
4.688E-01 & 1.801E-01 & 1.208E-01 & 1.462E-01 & 3.489E-03 & 3.861E-01 \\
5.000E-01 & 1.907E-01 & 1.220E-01 & 1.474E-01 & 2.910E-03 & 4.366E-01 \\
\end{tabular}
\end{table}

\begin{table}
\caption{$N=0.5$, $\kappa=0.3$}
\label{table n=0.5 k=0.3}
\begin{tabular}{llllll}
$\tilde{r}_A$ & $\tilde{J}$ & $\tilde{M}$ & $\tilde{M}_0$ & $\tilde{\Omega}^2$ & $f_x$\\\hline
1.000E-08 & 2.013E-01 & 1.556E-01 & 1.762E-01 & 1.429E-02 & 2.520E-01 \\
3.125E-02 & 2.013E-01 & 1.555E-01 & 1.762E-01 & 1.410E-02 & 2.536E-01 \\
6.250E-02 & 2.013E-01 & 1.554E-01 & 1.761E-01 & 1.359E-02 & 2.577E-01 \\
9.375E-02 & 2.015E-01 & 1.554E-01 & 1.762E-01 & 1.269E-02 & 2.656E-01 \\
1.250E-01 & 2.021E-01 & 1.554E-01 & 1.765E-01 & 1.159E-02 & 2.765E-01 \\
1.562E-01 & 2.031E-01 & 1.555E-01 & 1.768E-01 & 1.037E-02 & 2.911E-01 \\
1.875E-01 & 2.048E-01 & 1.556E-01 & 1.773E-01 & 9.128E-03 & 3.090E-01 \\
2.188E-01 & 2.069E-01 & 1.559E-01 & 1.778E-01 & 7.901E-03 & 3.308E-01 \\
2.500E-01 & 2.096E-01 & 1.562E-01 & 1.784E-01 & 6.760E-03 & 3.569E-01 \\
2.812E-01 & 2.127E-01 & 1.565E-01 & 1.790E-01 & 5.705E-03 & 3.878E-01 \\
3.125E-01 & 2.163E-01 & 1.568E-01 & 1.796E-01 & 4.770E-03 & 4.243E-01 \\
3.438E-01 & 2.203E-01 & 1.571E-01 & 1.801E-01 & 3.942E-03 & 4.673E-01 \\
3.750E-01 & 2.247E-01 & 1.574E-01 & 1.808E-01 & 3.220E-03 & 5.179E-01 \\
4.062E-01 & 2.298E-01 & 1.577E-01 & 1.814E-01 & 2.607E-03 & 5.773E-01 \\
4.375E-01 & 2.359E-01 & 1.582E-01 & 1.822E-01 & 2.096E-03 & 6.472E-01 \\
4.688E-01 & 2.413E-01 & 1.587E-01 & 1.830E-01 & 1.646E-03 & 7.308E-01 \\
5.000E-01 & 2.499E-01 & 1.594E-01 & 1.839E-01 & 1.308E-03 & 8.297E-01 \\
\end{tabular}
\end{table}

\begin{table}
\caption{$N=1.0$, $\kappa=0.3$}
\label{table n=1.0 k=0.3}
\begin{tabular}{llllll}
$\tilde{r}_A$ & $\tilde{J}$ & $\tilde{M}$ & $\tilde{M}_0$ & $\tilde{\Omega}^2$ & $f_x$\\\hline
1.000E-08 & 2.557E-01 & 1.882E-01 & 2.007E-01 & 6.089E-03 & 4.906E-01 \\
3.125E-02 & 2.594E-01 & 1.881E-01 & 1.999E-01 & 6.206E-03 & 4.901E-01 \\
6.250E-02 & 2.576E-01 & 1.882E-01 & 2.007E-01 & 5.827E-03 & 5.021E-01 \\
9.375E-02 & 2.589E-01 & 1.884E-01 & 2.011E-01 & 5.453E-03 & 5.174E-01 \\
1.250E-01 & 2.606E-01 & 1.886E-01 & 2.016E-01 & 4.984E-03 & 5.395E-01 \\
1.562E-01 & 2.630E-01 & 1.889E-01 & 2.022E-01 & 4.466E-03 & 5.681E-01 \\
1.875E-01 & 2.660E-01 & 1.893E-01 & 2.029E-01 & 3.930E-03 & 6.036E-01 \\
2.188E-01 & 2.697E-01 & 1.897E-01 & 2.036E-01 & 3.407E-03 & 6.457E-01 \\
2.500E-01 & 2.738E-01 & 1.902E-01 & 2.045E-01 & 2.908E-03 & 6.975E-01 \\
2.812E-01 & 2.784E-01 & 1.906E-01 & 2.053E-01 & 2.452E-03 & 7.583E-01 \\
3.125E-01 & 2.837E-01 & 1.911E-01 & 2.062E-01 & 2.044E-03 & 8.292E-01 \\
3.438E-01 & 2.897E-01 & 1.916E-01 & 2.071E-01 & 1.687E-03 & 9.117E-01 \\
3.750E-01 & 2.966E-01 & 1.922E-01 & 2.080E-01 & 1.379E-03 & 1.008E+00 \\
4.062E-01 & 3.041E-01 & 1.928E-01 & 2.089E-01 & 1.117E-03 & 1.120E+00 \\
4.375E-01 & 3.125E-01 & 1.934E-01 & 2.099E-01 & 8.958E-04 & 1.252E+00 \\
4.688E-01 & 3.214E-01 & 1.940E-01 & 2.109E-01 & 7.092E-04 & 1.408E+00 \\
5.000E-01 & 3.314E-01 & 1.947E-01 & 2.119E-01 & 5.562E-04 & 1.590E+00 \\
\end{tabular}
\end{table}

\begin{table}
\caption{$N=0.0$, $\kappa=0.5$ (Highly relativistic)}
\label{table n=0.0 k=0.5}
\begin{tabular}{llllll}
$\tilde{r}_A$ & $\tilde{J}$ & $\tilde{M}$ & $\tilde{M}_0$ & $\tilde{\Omega}^2$ & $f_x$\\\hline
1.000E-08 & 1.911E-01 & 1.523E-01 & 1.833E-01 & 2.866E-02 & 1.209E-01 \\
3.125E-02 & 1.901E-01 & 1.519E-01 & 1.831E-01 & 2.819E-02 & 1.214E-01 \\
6.250E-02 & 1.889E-01 & 1.516E-01 & 1.829E-01 & 2.749E-02 & 1.225E-01 \\
9.375E-02 & 1.880E-01 & 1.516E-01 & 1.836E-01 & 2.564E-02 & 1.263E-01 \\
1.250E-01 & 1.871E-01 & 1.514E-01 & 1.837E-01 & 2.362E-02 & 1.315E-01 \\
1.562E-01 & 1.867E-01 & 1.516E-01 & 1.846E-01 & 2.122E-02 & 1.385E-01 \\
1.875E-01 & 1.871E-01 & 1.518E-01 & 1.854E-01 & 1.875E-02 & 1.473E-01 \\
2.188E-01 & 1.883E-01 & 1.523E-01 & 1.865E-01 & 1.623E-02 & 1.586E-01 \\
2.500E-01 & 1.900E-01 & 1.528E-01 & 1.877E-01 & 1.396E-02 & 1.709E-01 \\
2.812E-01 & 1.922E-01 & 1.534E-01 & 1.890E-01 & 1.183E-02 & 1.866E-01 \\
3.125E-01 & 1.952E-01 & 1.538E-01 & 1.898E-01 & 9.972E-03 & 2.049E-01 \\
3.438E-01 & 1.976E-01 & 1.539E-01 & 1.903E-01 & 8.243E-03 & 2.270E-01 \\
3.750E-01 & 2.007E-01 & 1.539E-01 & 1.908E-01 & 6.785E-03 & 2.520E-01 \\
4.062E-01 & 2.029E-01 & 1.537E-01 & 1.912E-01 & 5.397E-03 & 2.835E-01 \\
4.375E-01 & 2.060E-01 & 1.541E-01 & 1.922E-01 & 4.280E-03 & 3.191E-01 \\
4.688E-01 & 2.081E-01 & 1.548E-01 & 1.940E-01 & 3.310E-03 & 3.618E-01 \\
5.000E-01 & 2.187E-01 & 1.559E-01 & 1.951E-01 & 2.703E-03 & 4.115E-01 \\
\end{tabular}
\end{table}

\begin{table}
\caption{$N=0.5$, $\kappa=0.5$ (Highly relativistic)}
\label{table n=0.5 k=0.5}
\begin{tabular}{llllll}
$\tilde{r}_A$ & $\tilde{J}$ & $\tilde{M}$ & $\tilde{M}_0$ & $\tilde{\Omega}^2$ & $f_x$\\\hline
1.000E-08 & 2.589E-01 & 1.956E-01 & 2.161E-01 & 1.217E-02 & 2.379E-01 \\
3.125E-02 & 2.586E-01 & 1.955E-01 & 2.161E-01 & 1.199E-02 & 2.394E-01 \\
6.250E-02 & 2.586E-01 & 1.956E-01 & 2.166E-01 & 1.154E-02 & 2.439E-01 \\
9.375E-02 & 2.590E-01 & 1.960E-01 & 2.175E-01 & 1.079E-02 & 2.524E-01 \\
1.250E-01 & 2.598E-01 & 1.965E-01 & 2.187E-01 & 9.869E-03 & 2.640E-01 \\
1.562E-01 & 2.613E-01 & 1.970E-01 & 2.200E-01 & 8.839E-03 & 2.791E-01 \\
1.875E-01 & 2.633E-01 & 1.977E-01 & 2.215E-01 & 7.779E-03 & 2.981E-01 \\
2.188E-01 & 2.660E-01 & 1.984E-01 & 2.232E-01 & 6.733E-03 & 3.209E-01 \\
2.500E-01 & 2.694E-01 & 1.992E-01 & 2.249E-01 & 5.750E-03 & 3.480E-01 \\
2.812E-01 & 2.728E-01 & 2.000E-01 & 2.266E-01 & 4.839E-03 & 3.805E-01 \\
3.125E-01 & 2.769E-01 & 2.007E-01 & 2.283E-01 & 4.029E-03 & 4.189E-01 \\
3.438E-01 & 2.814E-01 & 2.014E-01 & 2.299E-01 & 3.313E-03 & 4.641E-01 \\
3.750E-01 & 2.861E-01 & 2.022E-01 & 2.317E-01 & 2.688E-03 & 5.173E-01 \\
4.062E-01 & 2.920E-01 & 2.029E-01 & 2.333E-01 & 2.164E-03 & 5.790E-01 \\
4.375E-01 & 2.990E-01 & 2.038E-01 & 2.350E-01 & 1.730E-03 & 6.517E-01 \\
4.688E-01 & 3.051E-01 & 2.045E-01 & 2.367E-01 & 1.350E-03 & 7.385E-01 \\
5.000E-01 & 3.141E-01 & 2.055E-01 & 2.385E-01 & 1.058E-03 & 8.408E-01 \\
\end{tabular}
\end{table}

\begin{table}
\caption{$N=1.0$, $\kappa=0.5$ (Highly relativistic)}
\label{table n=1.0 k=0.5}
\begin{tabular}{llllll}
$\tilde{r}_A$ & $\tilde{J}$ & $\tilde{M}$ & $\tilde{M}_0$ & $\tilde{\Omega}^2$ & $f_x$\\\hline
1.000E-08 & 3.578E-01 & 2.399E-01 & 2.437E-01 & 4.250E-03 & 5.513E-01 \\
3.125E-02 & 3.583E-01 & 2.403E-01 & 2.440E-01 & 4.206E-03 & 5.555E-01 \\
6.250E-02 & 3.592E-01 & 2.406E-01 & 2.446E-01 & 4.048E-03 & 5.664E-01 \\
9.375E-02 & 3.608E-01 & 2.410E-01 & 2.455E-01 & 3.800E-03 & 5.844E-01 \\
1.250E-01 & 3.637E-01 & 2.415E-01 & 2.465E-01 & 3.489E-03 & 6.089E-01 \\
1.562E-01 & 3.667E-01 & 2.422E-01 & 2.479E-01 & 3.136E-03 & 6.422E-01 \\
1.875E-01 & 3.711E-01 & 2.430E-01 & 2.494E-01 & 2.769E-03 & 6.824E-01 \\
2.188E-01 & 3.757E-01 & 2.438E-01 & 2.511E-01 & 2.402E-03 & 7.325E-01 \\
2.500E-01 & 3.816E-01 & 2.446E-01 & 2.528E-01 & 2.054E-03 & 7.906E-01 \\
2.812E-01 & 3.878E-01 & 2.455E-01 & 2.546E-01 & 1.730E-03 & 8.601E-01 \\
3.125E-01 & 3.927E-01 & 2.466E-01 & 2.569E-01 & 1.424E-03 & 9.449E-01 \\
3.438E-01 & 4.025E-01 & 2.473E-01 & 2.582E-01 & 1.185E-03 & 1.037E+00 \\
3.750E-01 & 4.120E-01 & 2.481E-01 & 2.598E-01 & 9.668E-04 & 1.145E+00 \\
4.062E-01 & 4.232E-01 & 2.489E-01 & 2.613E-01 & 7.820E-04 & 1.268E+00 \\
4.375E-01 & 4.357E-01 & 2.498E-01 & 2.628E-01 & 6.269E-04 & 1.410E+00 \\
4.688E-01 & 4.489E-01 & 2.507E-01 & 2.644E-01 & 4.971E-04 & 1.577E+00 \\
5.000E-01 & 4.652E-01 & 2.517E-01 & 2.659E-01 & 3.926E-04 & 1.767E+00 \\
\end{tabular}
\end{table}

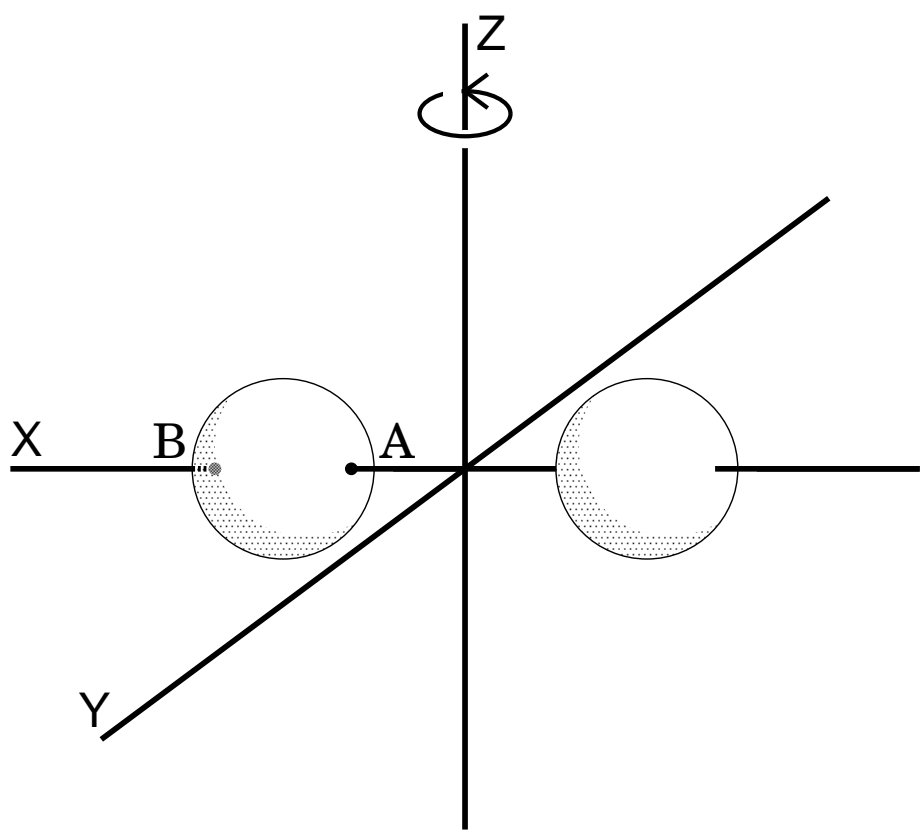
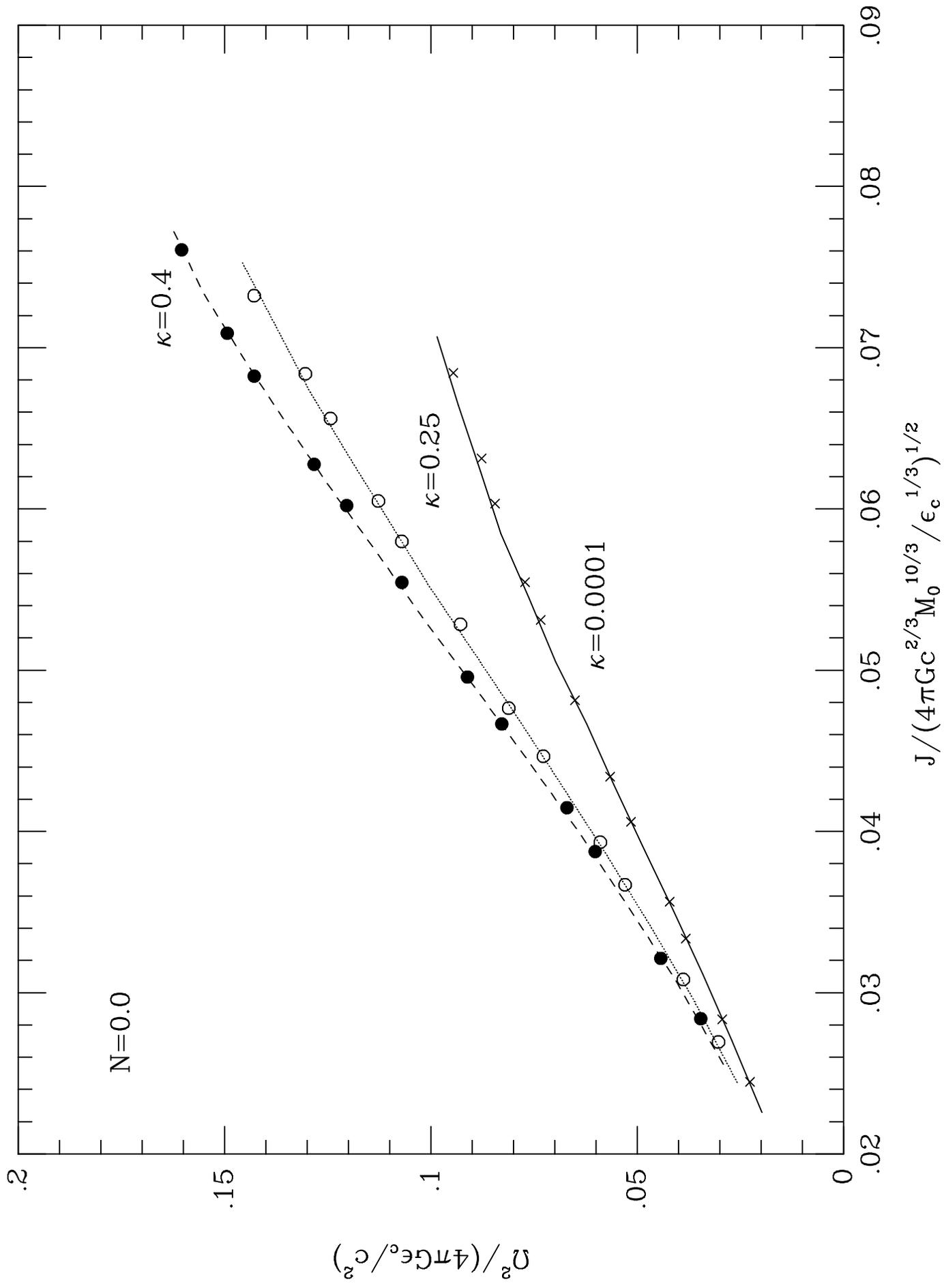
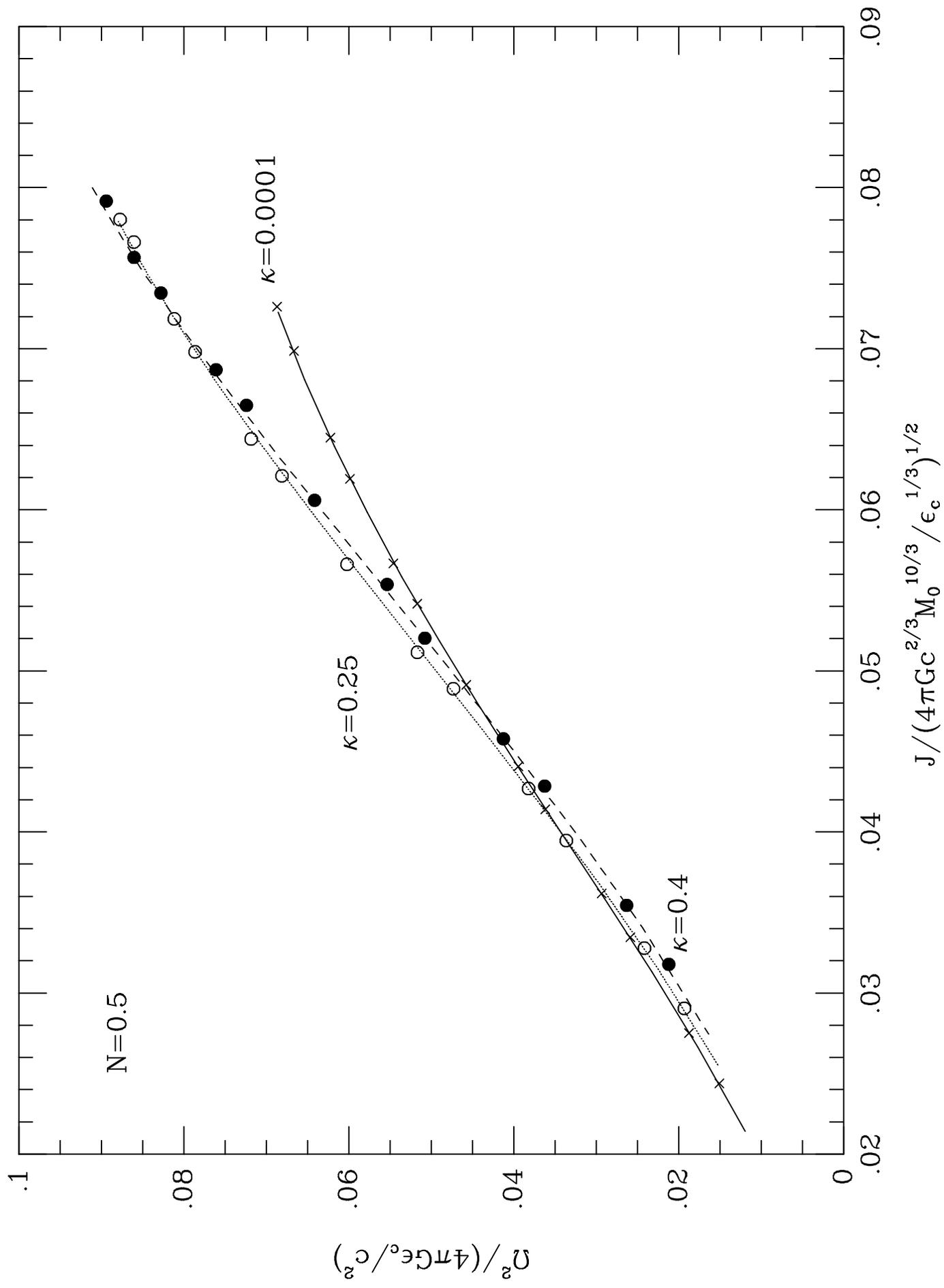
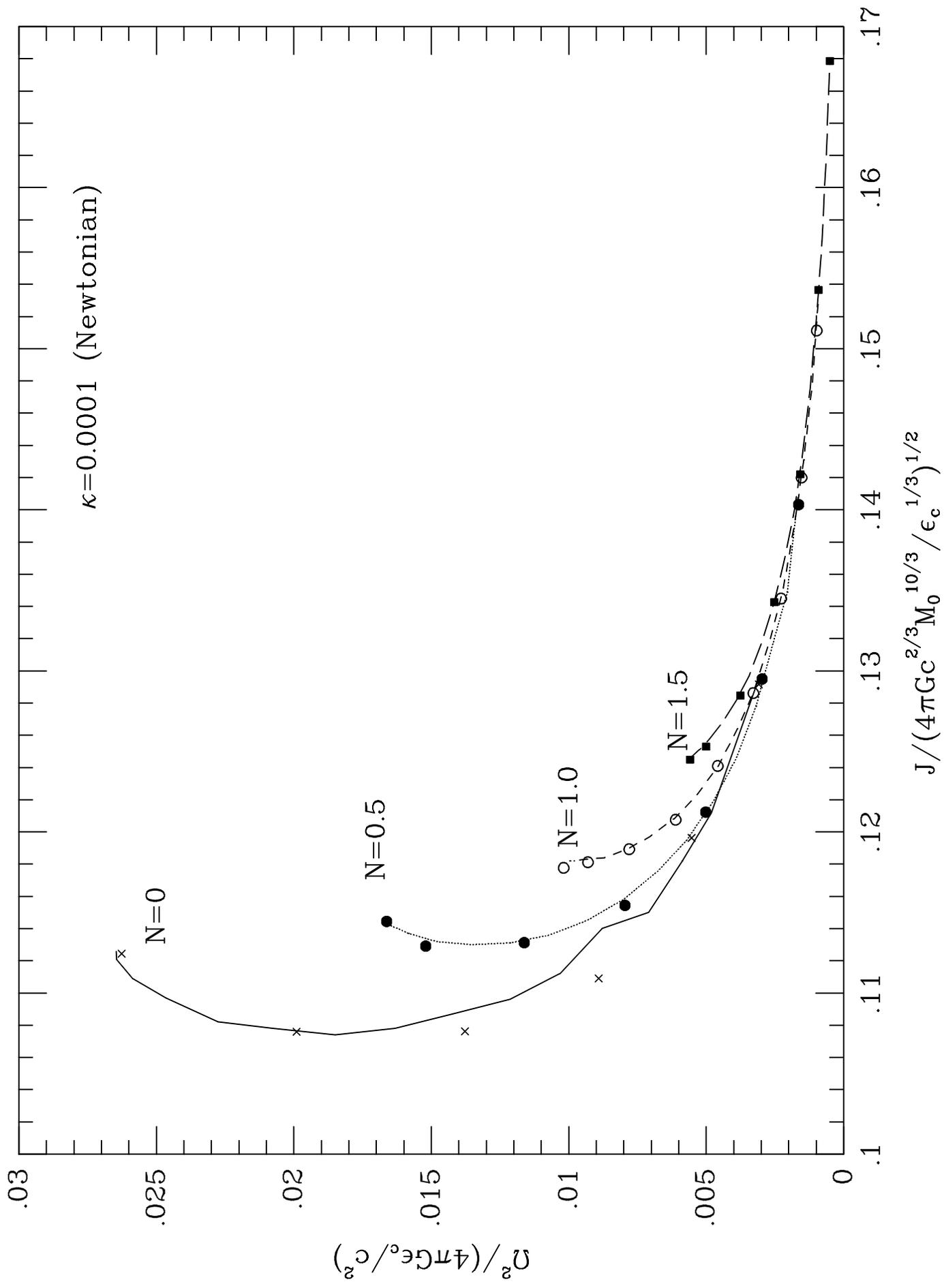
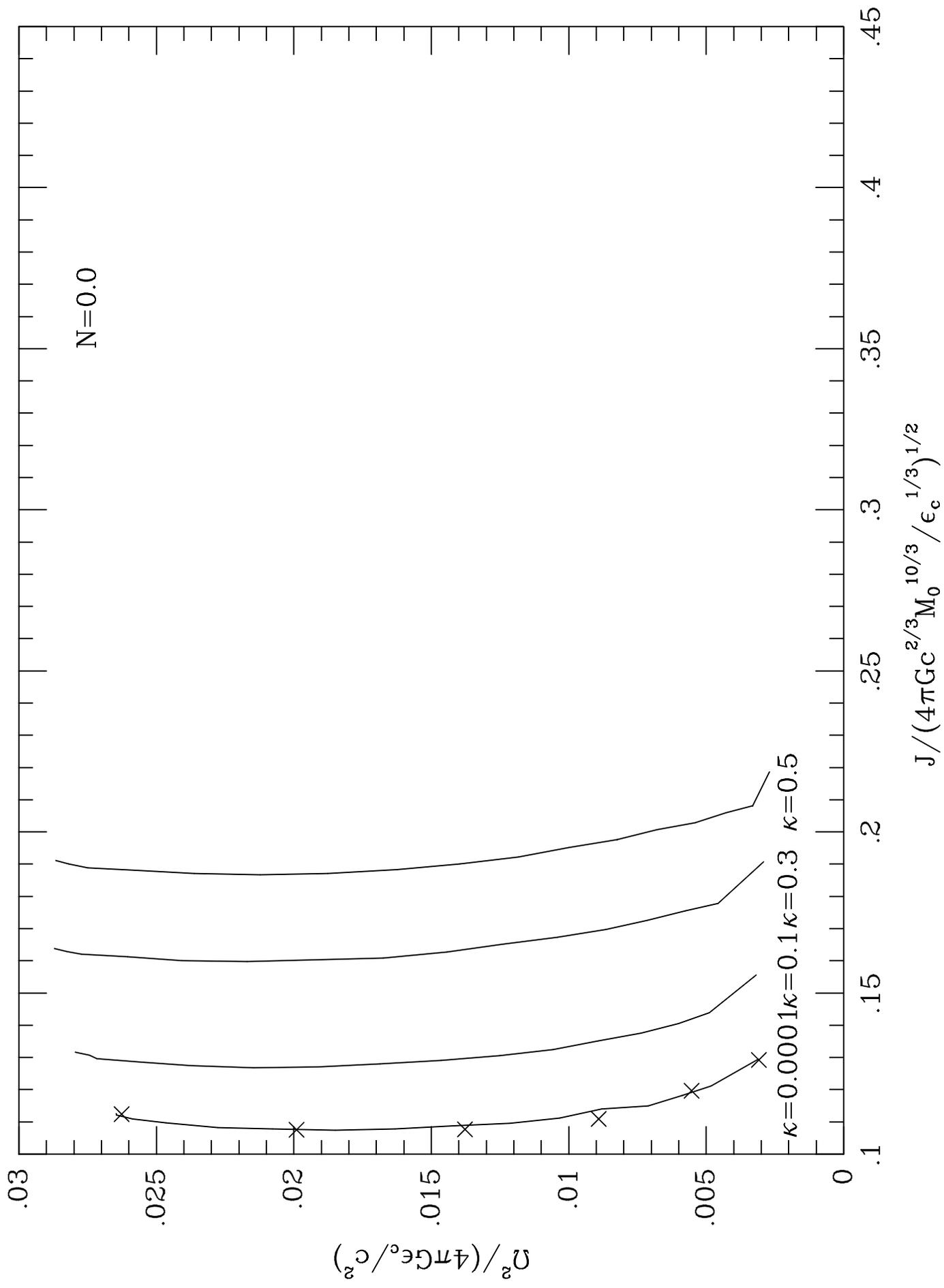
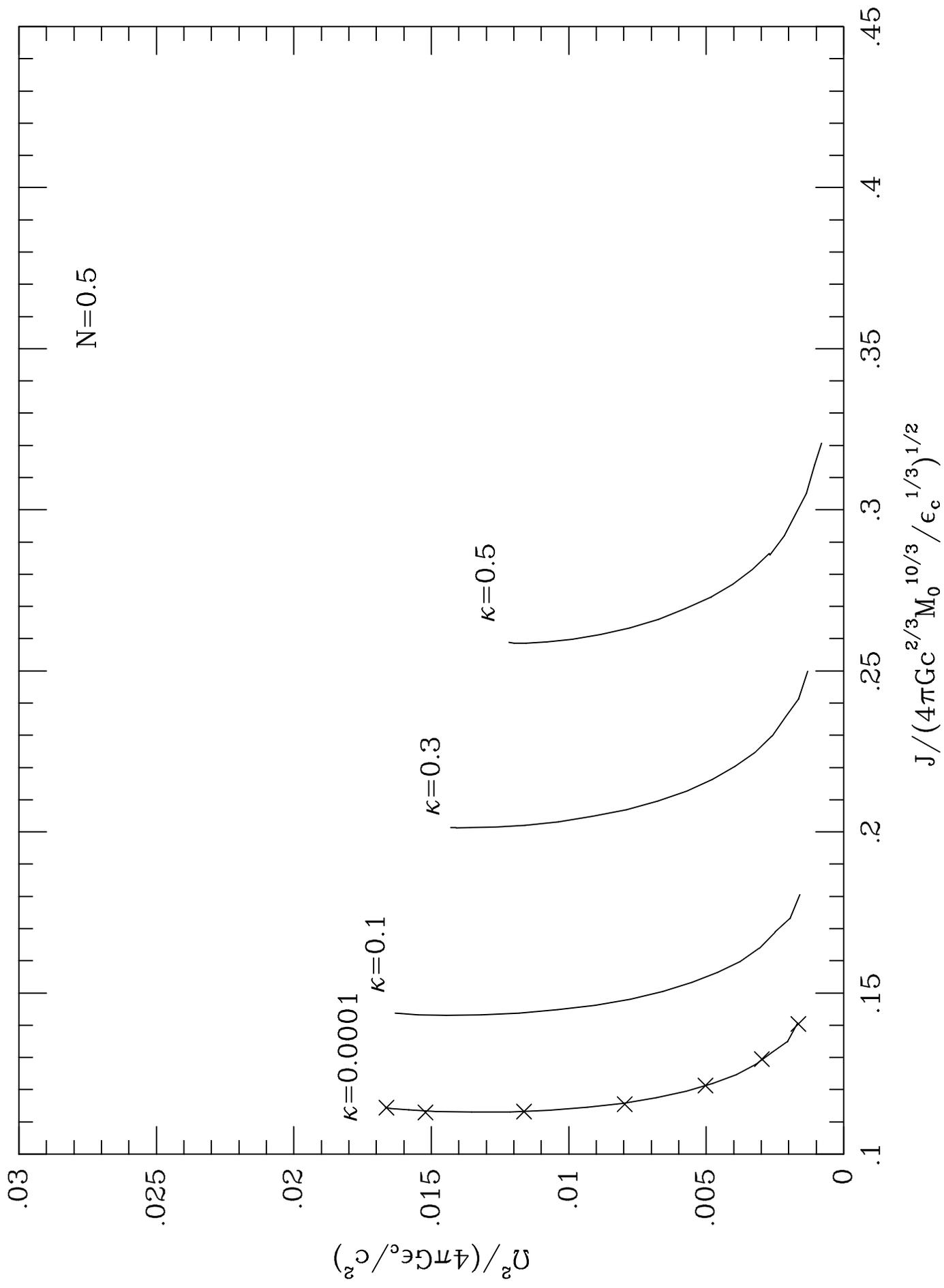
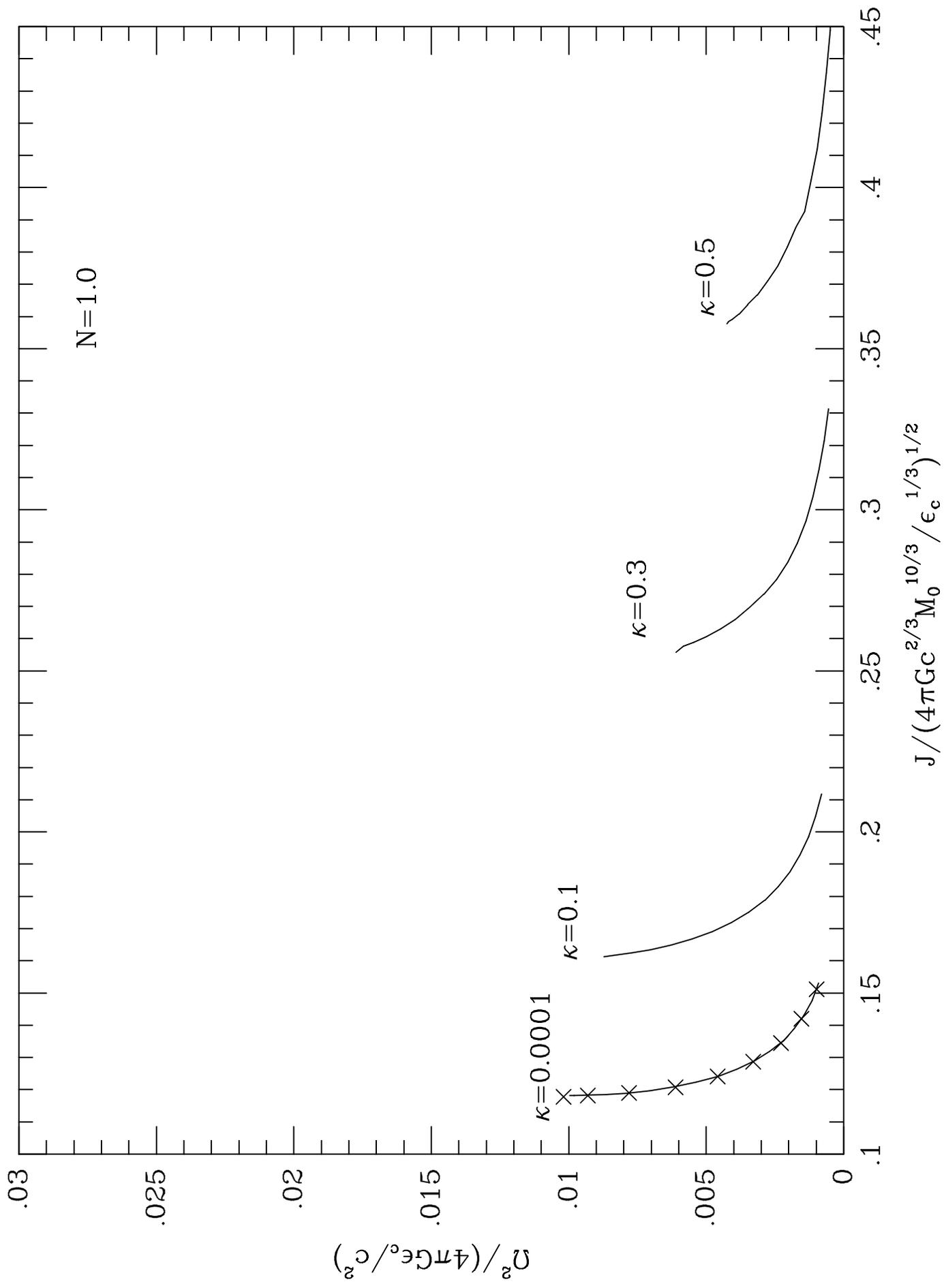
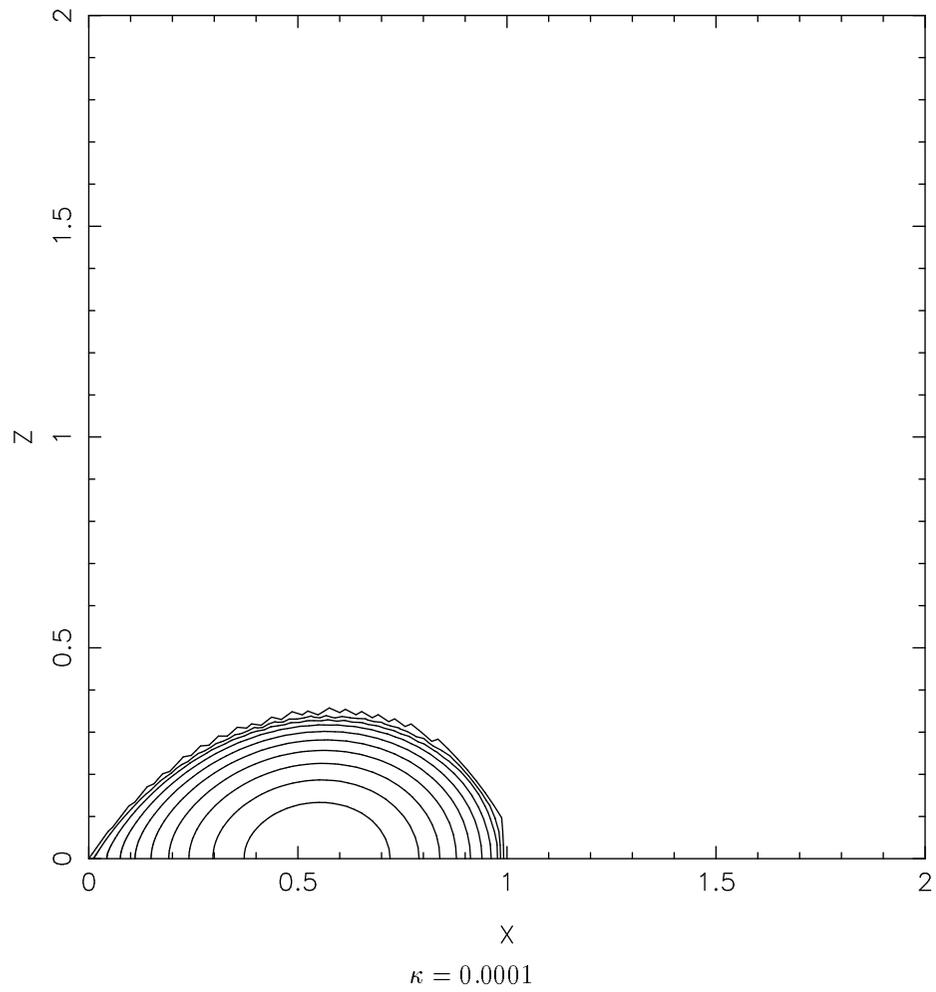
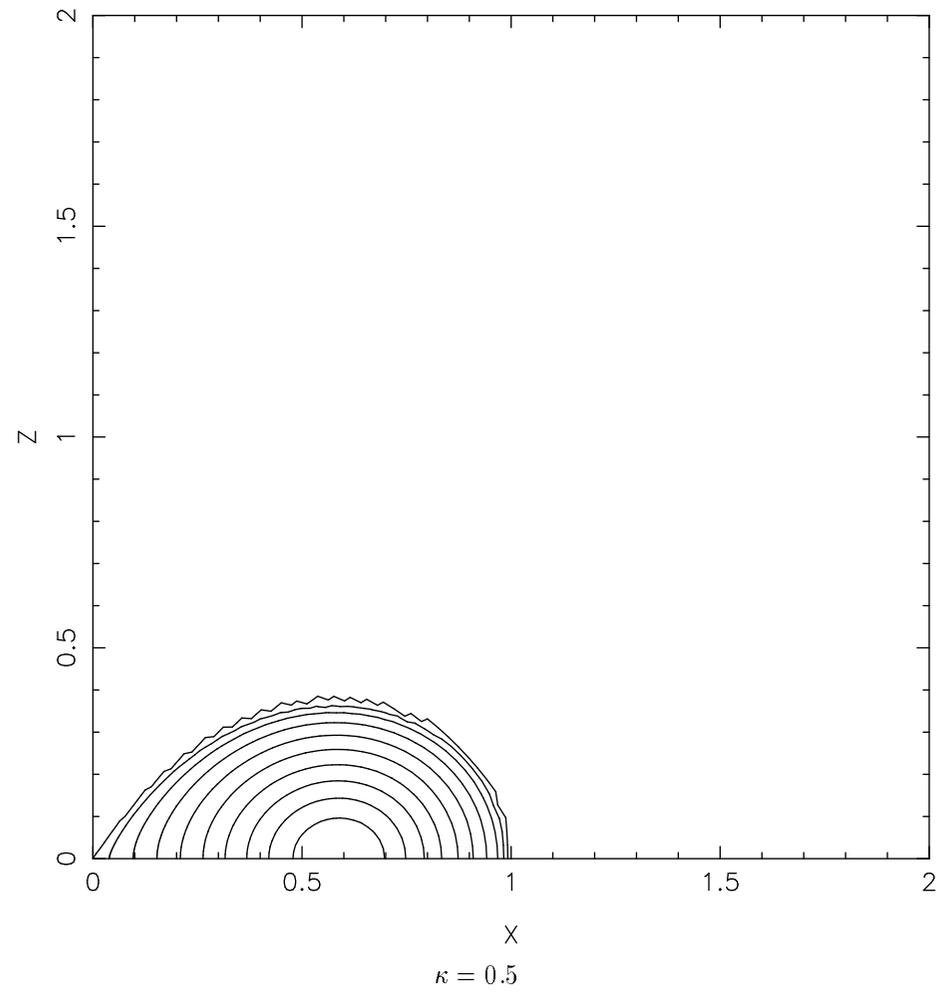
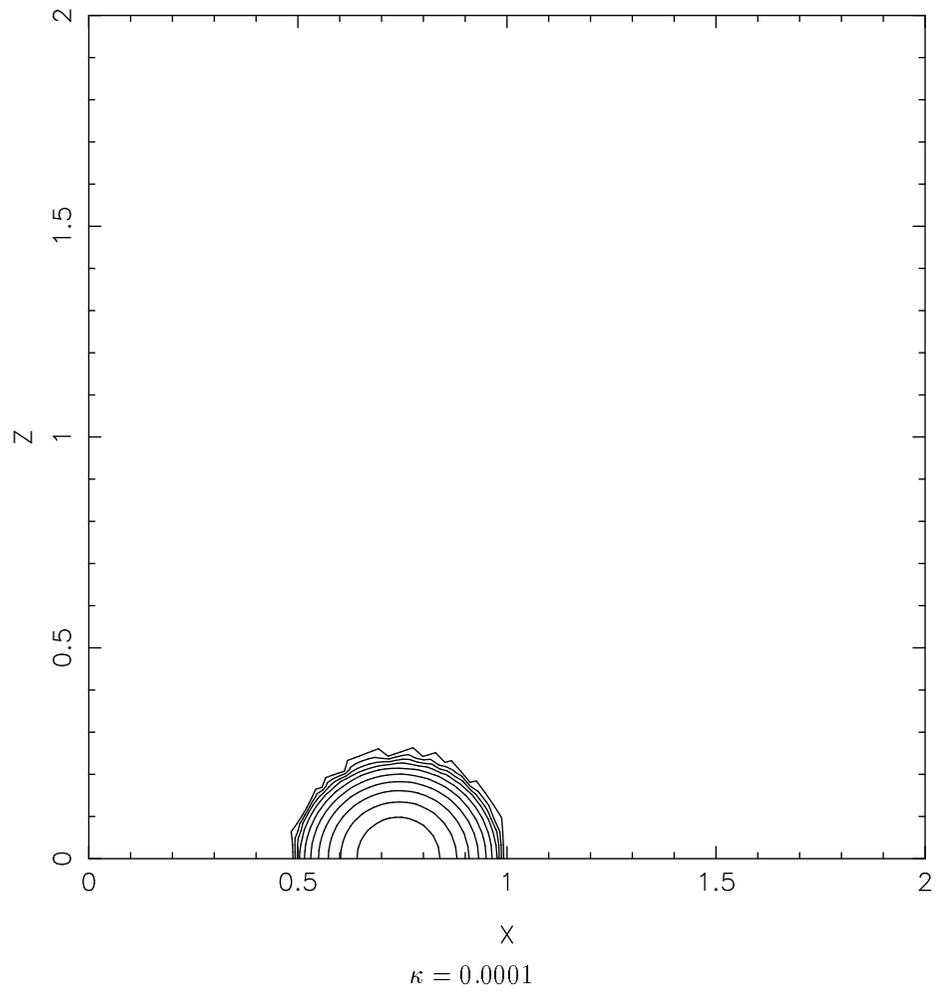
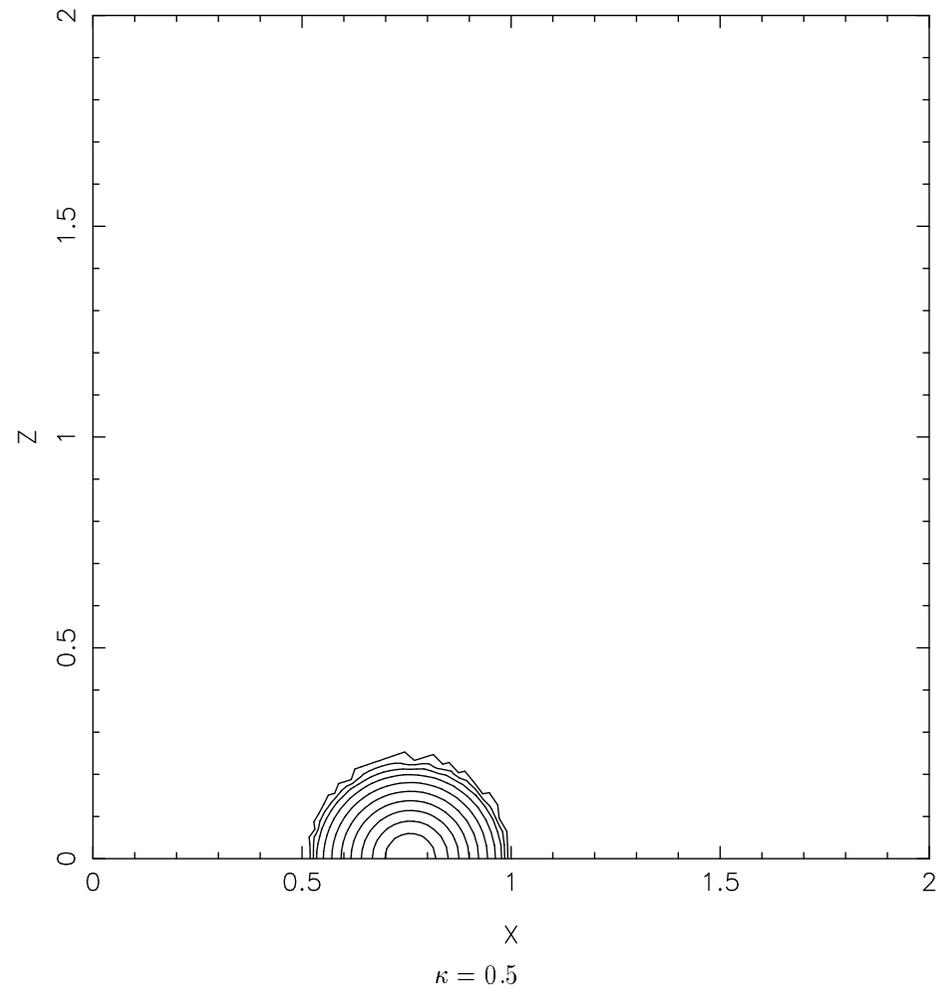
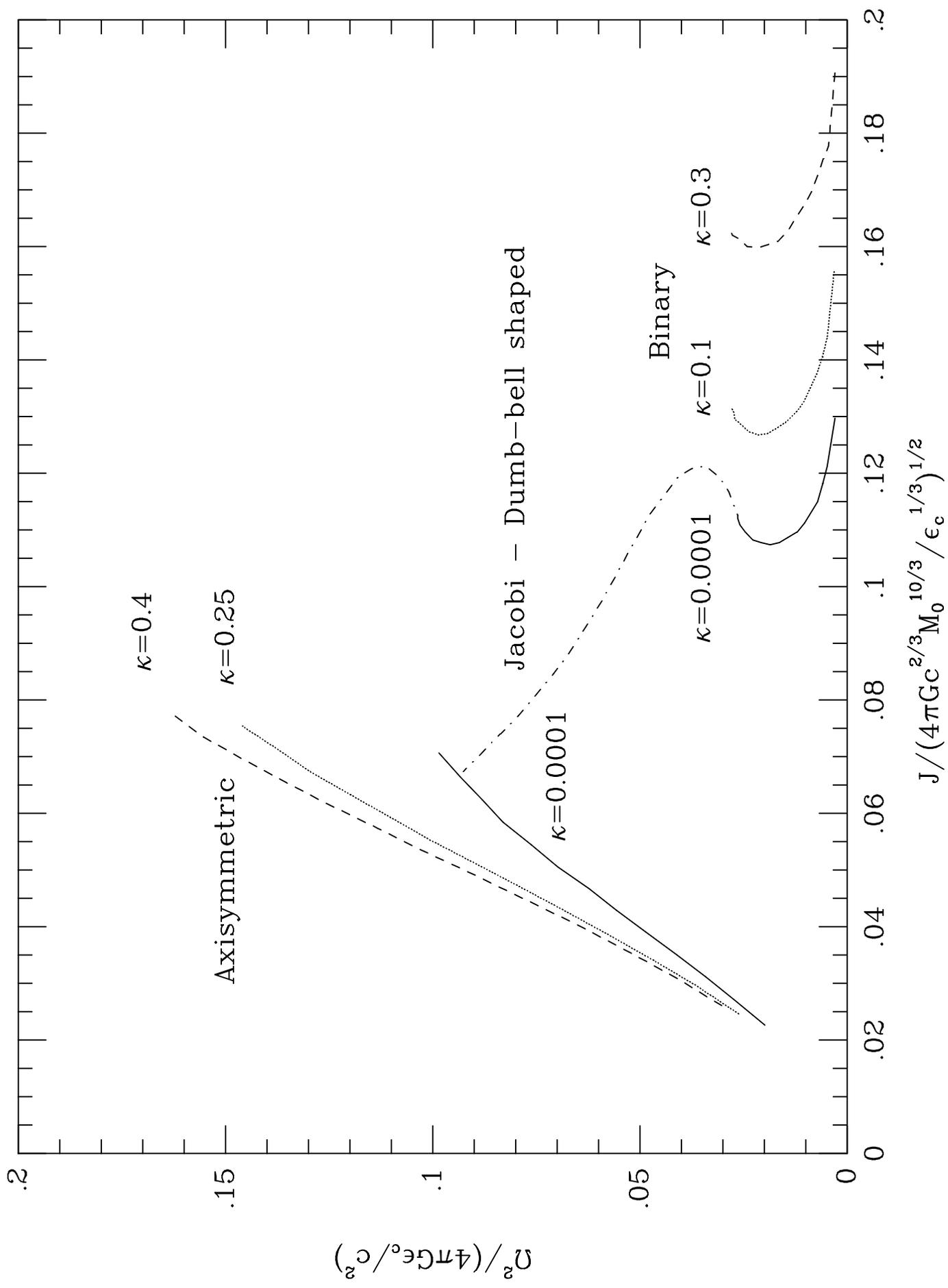
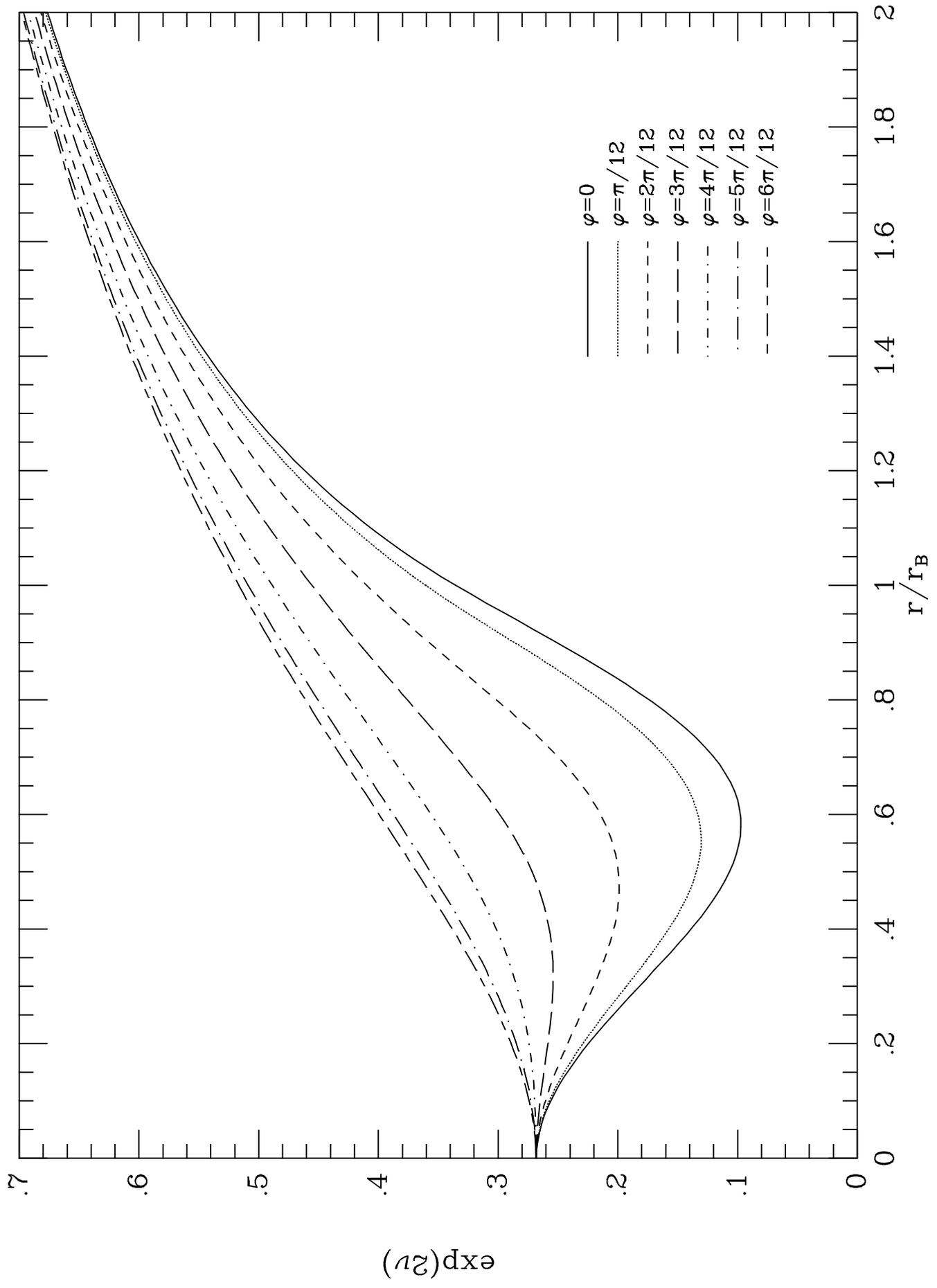
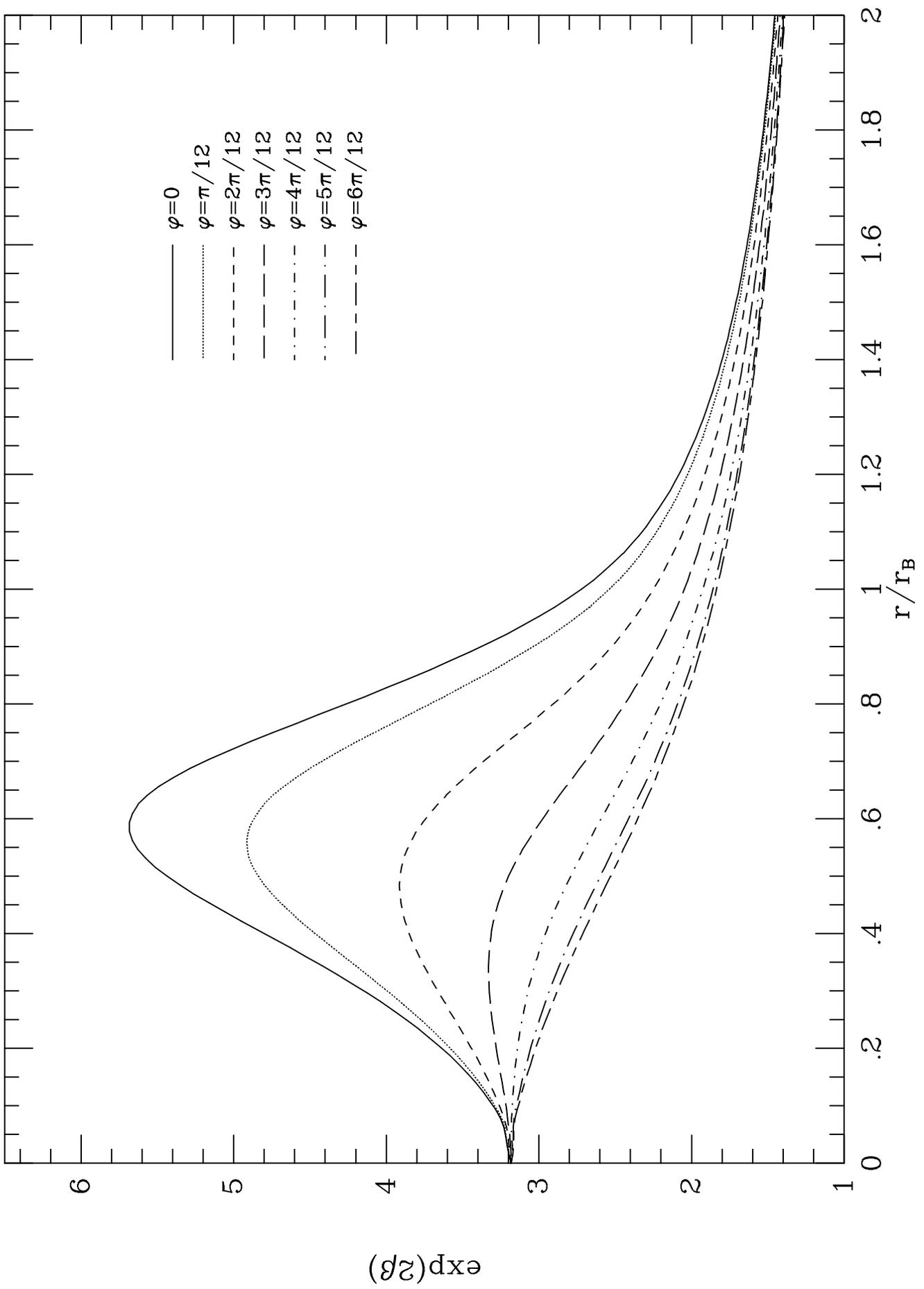
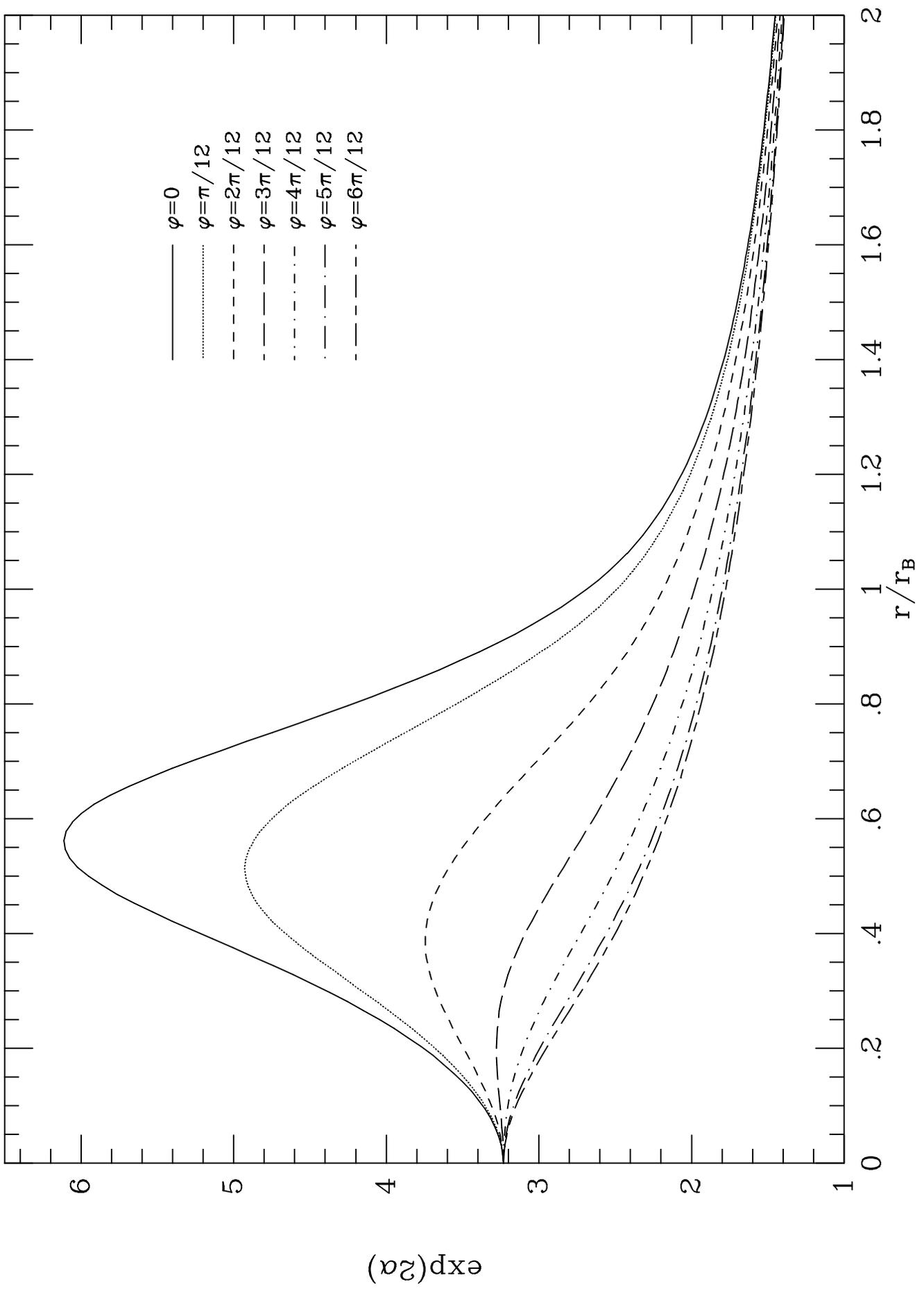
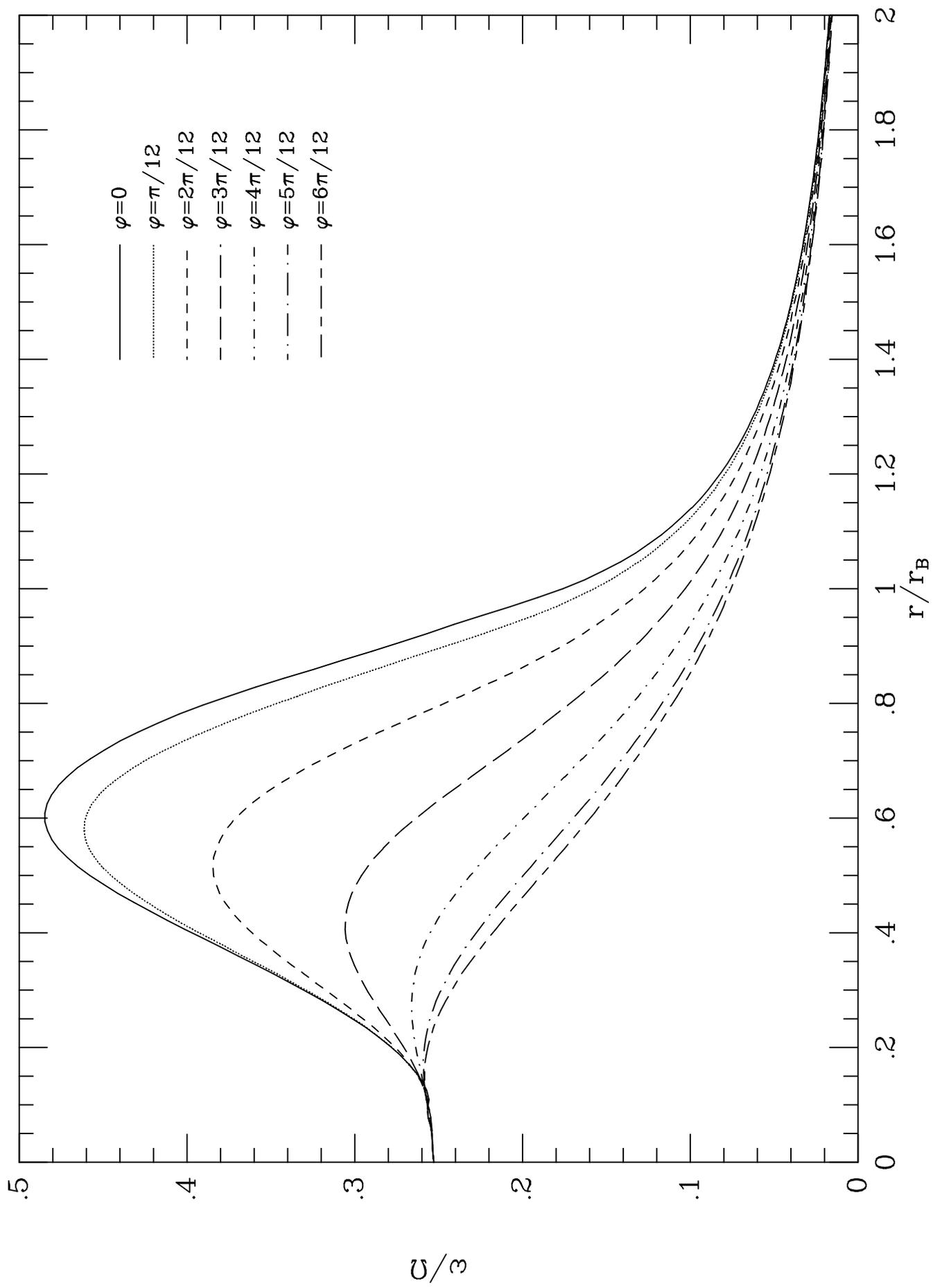
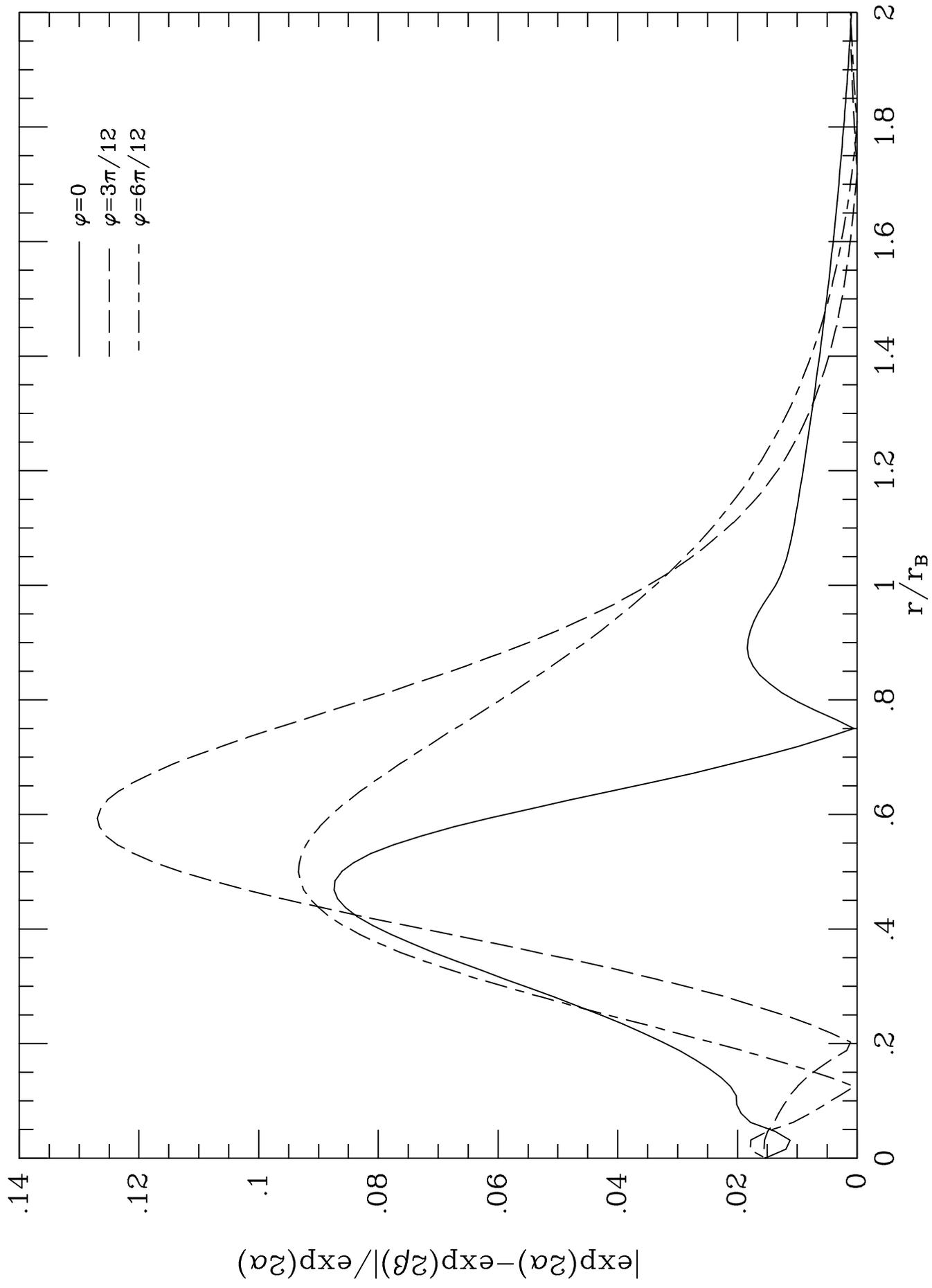
\end{document}